\documentclass[twoside]{article}

\usepackage{arxiv}
\usepackage{amsmath}

\usepackage{tikz} % Fig 15 (diffractive lightguide) uses this

\usepackage[utf8]{inputenc} % allow utf-8 input
\usepackage[T1]{fontenc}    % use 8-bit T1 fonts
\usepackage{hyperref}       % hyperlinks
\usepackage{url}            % simple URL typesetting
\usepackage{booktabs}       % professional-quality tables
\usepackage{amsfonts}       % blackboard math symbols
\usepackage{nicefrac}       % compact symbols for 1/2, etc.
\usepackage{microtype}      % microtypography
\usepackage{lipsum}
\usepackage{fancyhdr}       % header
\usepackage{graphicx}       % graphics
\graphicspath{{images/}}     % organize your images and other figures under media/ folder
\usepackage[sort]{cite}
\title{Positioning Monocular Optical See Through Head Worn Displays in Glasses for Everyday Wear 
}

\author{
    \textbf{Parth Arora}$^{1}$ \quad
    \textbf{Ethan Kimmel}$^{1}$ \quad
    \textbf{Katherine Huang}$^{1}$ \quad
    \textbf{Tyler Kwok}$^{1}$ \quad
    \textbf{Yukun Song}$^{1}$ \\
    \textbf{Sofia Vempala}$^{1}$ \quad
    \textbf{Georgianna Lin}$^{2}$ \quad
    \textbf{Ozan Cakmakci}$^{3}$ \quad
    \textbf{Thad Starner}$^{1, 3}$ \\
    \\
    $^{1}$Georgia Institute of Technology\quad
    $^{2}$University of Toronto\quad
    $^{3}$Google
}

\begin{document}
\maketitle

\begin{abstract}
Head-worn displays for everyday wear in the form of regular eyeglasses are technically feasible with recent advances in waveguide technology. One major design decision is determining where in the user's visual field to position the display. Centering the display in the principal point of gaze (PPOG) allows the user to switch attentional focus between the virtual and real images quickly, and best performance often occurs when the display is centered in PPOG or is centered vertically below PPOG. However, these positions are often undesirable in that they are considered interruptive or are associated with negative social perceptions by users. Offsetting the virtual image may be preferred when tasks involve driving, walking, or social interaction. This paper consolidates findings from recent studies on monocular optical see-through HWDs (OST-HWDs), focusing on potential for interruption, comfort, performance, and social perception. 
For text-based tasks, which serve as a proxy for many monocular OST-HWD tasks, we recommend a 15° horizontal field of view (FOV) with the virtual image in the right lens vertically centered but offset to +8.7° to +23.7° toward the ear. Glanceable content can be offset up to +30° for short interactions. 
% For the task of captioning conversations for users who are hard of hearing, which serves as a proxy for many monocular OST-HWD tasks, we recommend a 15° horizontal field of view (FOV) with the virtual image vertically centered but offset to +8.7° to +23.7° toward the ear and upto +30° for shorter interactions.

\end{abstract}


\section{Introduction}
\begin{figure}[ht]
    \centering
    \includegraphics[width=0.8\linewidth]{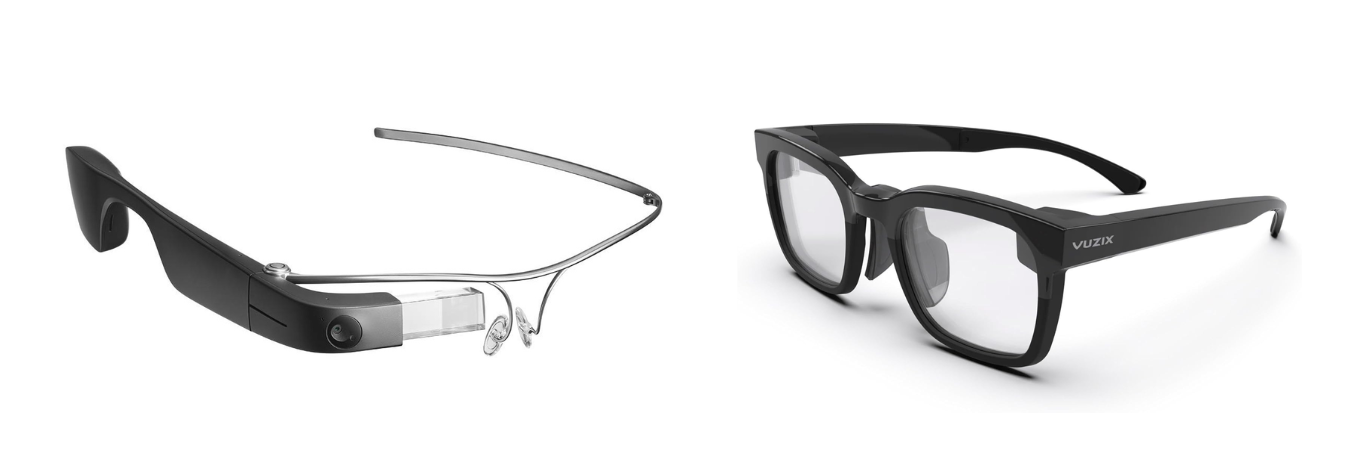}
    \caption{Google Glass in 2014 and Vuzix Z100 in 2024.}
    \label{fig:lightGlasses}
\end{figure}

Head-worn displays (HWDs)       \cite      {jain2018towards, miller2017use, schipper2017caption, velger1998helmet} are wearable virtual displays integrated into eyeglasses or head mounts that present data directly within the user's visual field       \cite      {kim2016augmented}. They encompass a diverse range of devices, from bulky headsets like the Magic Leap       \cite      {MagicLeap}, HoloLens       \cite      {HoloLens}, and Apple Vision Pro       \cite      {AppleVisionPro} to glasses-like devices such as Meta Orion       \cite      {MetaOrion} and Snap Spectacles       \cite      {SnapSpectacles}, which resemble regular eyewear although still bulkier than standard glasses. They also include devices like Google Glass, launched over a decade ago, which introduced mainstream consumer monocular optical see-through head-worn displays. These devices, often referred to as peripheral HWDs       \cite      {matthies2015properties}, feature a see-through display positioned in front of one eye. They can present a wide range of information, including video footage, text notifications       \cite      {kim2018impacts}, micro-interactions       \cite      {ashbrook2010enabling, starner2003enigmatic, starner2013google}, and contextually relevant suggestions from intelligent agents       \cite      {singletary2001learning, starner2002wearable}. Recent advancements in waveguide technologies have spurred a resurgence of these devices. This growth has sparked interest in exploring how HWDs can seamlessly integrate into daily activities       \cite      {aromaa2020awareness}. Smaller companies, such as Vuzix with the Z100 (Figure \ref{fig:lightGlasses}) and Even Realities with the G1, introduced lightweight HWDs for everyday wear weighing less than 40g in 2024       \cite      {Vuzix, EvenRealities}, demonstrating the technical feasibility of these devices and paving the way for more products to be introduced in 2025.  

Designing HWDs involves balancing a complex set of parameters within the constraints of manufacturing resources \cite{Peli_1999, francis2010cognitive}. One critical design decision is determining where the HWD’s virtual image should appear in the user’s visual field. 

%\subsection{Eye Movement}

Virtual image position affects eye movement during use, as some positions are more comfortable than others due to eye motor functions and habitual behavior       \cite      {boghen1974velocity}. Gaze shifts extending over $25^\circ$ tend to involve both head motion and eye movement       \cite      {freedman2008coordination, goossens1997human, land2009looking}, as opposed to just eye movement.  Eyes prefer to stay within a comfortable range (approximately 15 degrees left and right of center) relative to the head       \cite      {Peli_1999}. Hence, there is a range of eye motion beyond which it may become uncomfortable to consume content. Positioning the content in a central position would make sense intuitively. It also allows the user to switch attentional focus between the virtual and real images quickly.

%\subsection{Interruption}

However, positioning the display at the center of the user’s visual field can sometimes be interruptive. Such disruption is particularly true when content is displayed without being explicitly triggered, such as notifications or proactive AI suggestions. Research on head-up displays (HUDs) in aviation has identified a phenomenon called cognitive capture. In this condition, users focus on virtual information rather than the physical world, and their attention is tunneled into the virtual realm       \cite      {dowell}. Studies on HUD use while driving or walking have examined how interruptions, such as sudden notifications or system errors, can be uncomfortable for users. Shifting the display from the user's point of gaze (PPOG) can reduce the likelihood of cognitive capture and interruptions       \cite      {Foyle, dowell, yoo1999display, watanabe1999effect, tsimhoni2000display, tsimhoni2001detecting, mosur2024stepping}. Hence, positioning the image too centrally to the PPOG could be interruptive for a device designed for all day wear. 

%\subsection{Rivalry}

Another effect that could cause discomfort in monocular HWDs is \textbf{binocular rivalry}, a phenomenon that occurs when each eye is presented with a different image       \cite      {bayle2019binocular, velger1998helmet, blake1997can, patterson2007binocular}. When the brain receives conflicting signals from both eyes, dominance alternates between the two, leading to an unstable visual state. For a period, one eye’s image suppresses the other before the roles reverse. This continuous alternation can cause visual confusion. Laboratory studies have shown that high binocular rivalry conditions significantly reduce the efficiency of participants when quickly referring to lists of text at the PPOG        \cite      {laramee2002rivalry}. However, the common use of Google Glass (2012-2023) in users' everyday lives suggests that the effect may not be as worrisome as expected, at least when the image is above the user's PPOG. Furthermore, laboratory studies with Glass showed that the difference in efficiency between transparent and opaque versions of the display was small (-3\%), which also supports the hypothesis that binocular rivalry is less of a concern if the display is outside of PPOG  \cite       {guo2015order}.
Furthermore, Hershberger and Guerin’s research showed that placing displays below the PPOG significantly reduced binocular rivalry       \cite      {hershberger1975binocular},  and Peli and Hakkinen recommend positioning displays outside the user’s direct line of sight       \cite      {peli1990visual, hakkinen200461}, 

%\subsection{Task and social perception}

Additional factors, such as a user's performance on tasks supported by HWDs, are influenced by display position. If content is too central, it may interfere with tasks by being constantly in the way. Conversely, if it is positioned too far away, accessing information can become slower and less comfortable. Social perception is another critical factor. For instance, if the wearer is constantly looking over a conversational partner's head to read the display, the partner may have the impression that the wearer is less engaged with the conversation, much in the way that looking at one's watch may unintentionally signal impatience to a conversational partner.  
%bystanders' impressions of the user may deter the adoption of such devices. %With Google Glass       \cite      {GoogleGlass}, for example, notifications that required users to look up gave the impression that they were distracted from conversations.

Some software techniques can also improve user comfort. For example, Orlosky et al.       \cite      {orlosky2014managing, rzayev2020effects} found that users prefer having text in passive backgrounds and positions not blocking the main action in the scene. Furthermore, using gaze to display information in the user's periphery relative to their line of sight can improve usability \cite{yoshioPeripheral}. However, for devices to achieve this, they require a high field of view (FOV) at a minimum, and for specific implementations, power-intensive real-time gaze tracking and/or scene understanding.

\section{Testing with text}

Text-based content is one of the most demanding and representative benchmarks for evaluating comfort and usability in head-worn displays (HWDs). Unlike images or videos, reading requires sustained attention, increased cognitive load, and precise eye fixation, making it a rigorous benchmark for real-world usability. In addition, the primary interactions on current mobile devices—messaging, search, notifications, AI assistants, and microinteractions—are overwhelmingly text-based and are frequently pitched as key use cases for monocular, small-FOV HWDs. Their viability depends on how well they handle text-heavy tasks. Furthermore, \textbf{captioning on HWDs}       \cite      {olwal2020wearable, jones2014head, jain2018towards, miller2017use, schipper2017caption, velger1998helmet} is a critical example demonstrating the impact of sustained eye fixation. For deaf and hard-of-hearing (DHH) users, having to split attention between text and the environment could adversely impact both comfort and social perception. This difficulty underscores why researchers prioritize text-based tasks—if an HWD can make text consumption seamless, its advantageous properties can often be easily transferred to other less demanding applications. Thus, testing with text is not just a convenient choice—it is a necessity for ensuring these devices are comfortable and effective across their most essential use cases. 
%Thus, many studies rely on text-based tasks to evaluate their performance.

Striking a balance between ensuring noticeability and avoiding interruption is a fundamental challenge in notification system design       \cite      {mccrickard2003model}. Finding the "sweet spot" for display positioning is critical. The sections below summarize research on optimizing display positions to identify this ideal spot.

%       \cite      {starner1999everyday},      \cite      {toozkit} 

\section{Key Concepts}

\subsection{Field of View (FOV)}
In head worn displays (HWD), Field of View (FOV) refers to the angular extent of the virtual environment that can be seen through the display at any moment. FOV is a critical specification for devices such as AR/VR headsets because it directly impacts user immersion and the sense of presence. 
A higher FOV means the user can view a broader area of the virtual world without moving their head.
%, creating a more natural and engaging experience. 
For monocular HWDs, where only one eye is involved in viewing, the FOV is generally narrower compared to binocular systems as immersion and presence are of lesser concern.

\subsection{Field of Regard}
In HWDs, Field of Regard (FOR) refers to the total visual space a user can access, accounting for head movements, eye tracking, and device adjustments. While the FOV defines the immediate visual area within the device's viewport, the FOR includes the entire environment a user can perceive through physical actions like turning their head or shifting their gaze. FOR is particularly important in AR and mixed reality applications, where users interact with a digital overlay that spans their entire surroundings. A larger FOR can lead to a more immersive,  interactive, or efficient experience, as users can engage with digital content across a broader area. For example, interface designers for HWDs can construct a work environment in a cylinder around the user's head, allowing the user to place virtual windows (such as a calendar, editor, browser) in known fixed places relative to the user's head orientation such that the user can orient to different tasks quickly.

%, the FOR may be limited by head movement range or device orientation, as the single-eye setup inherently restricts peripheral vision. Still, users can navigate the virtual space through head movements, though the experience may be more confining compared to binocular systems.

\begin{figure}[ht]
    \centering
    \includegraphics[width=0.8\linewidth]{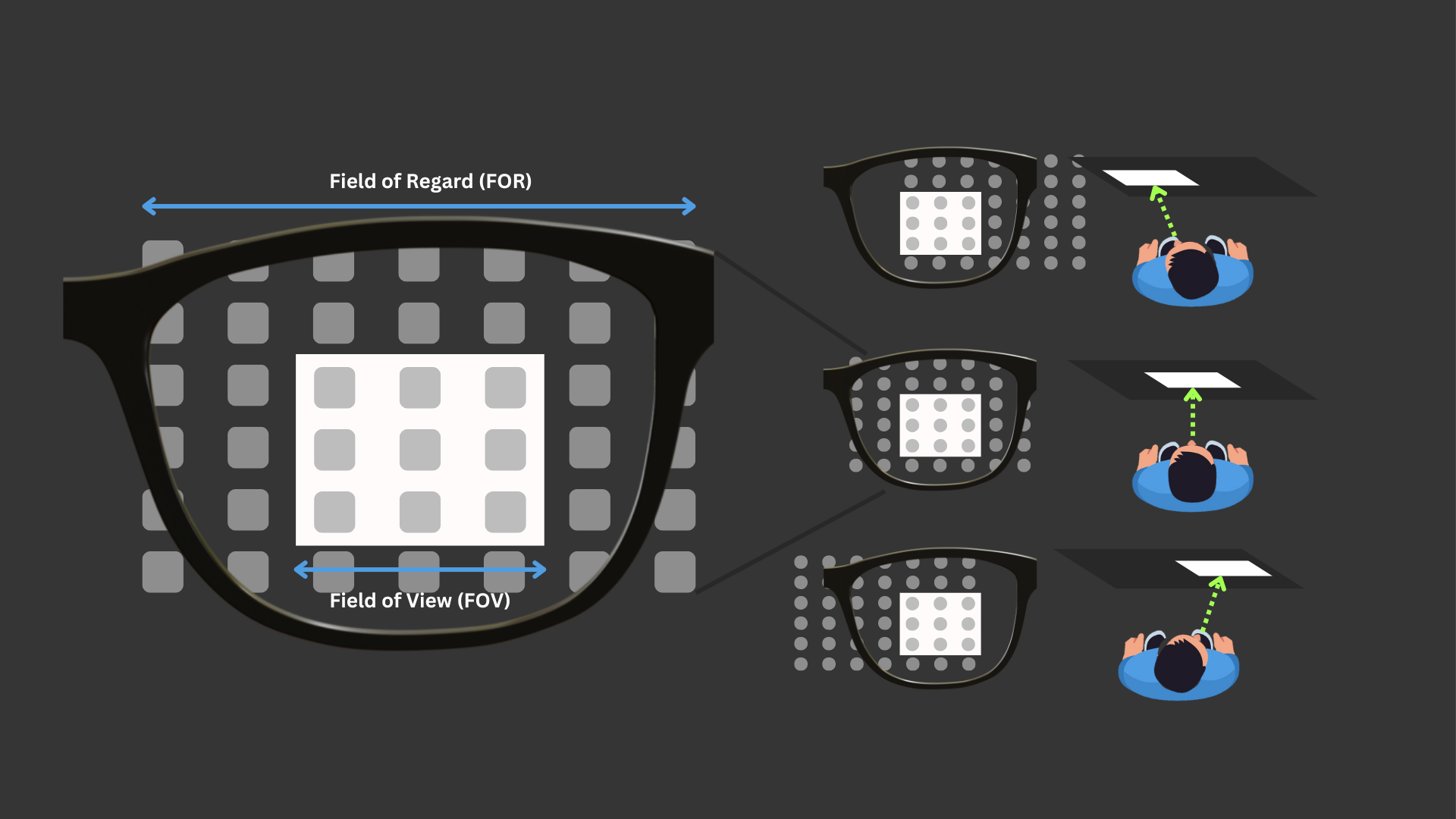}
    \caption{ Field of view versus field of regard.}
    \label{fig:for}
\end{figure}

\subsection{Visual Angles}
\begin{figure}[ht]
    \centering
    \includegraphics[width=0.5\linewidth]{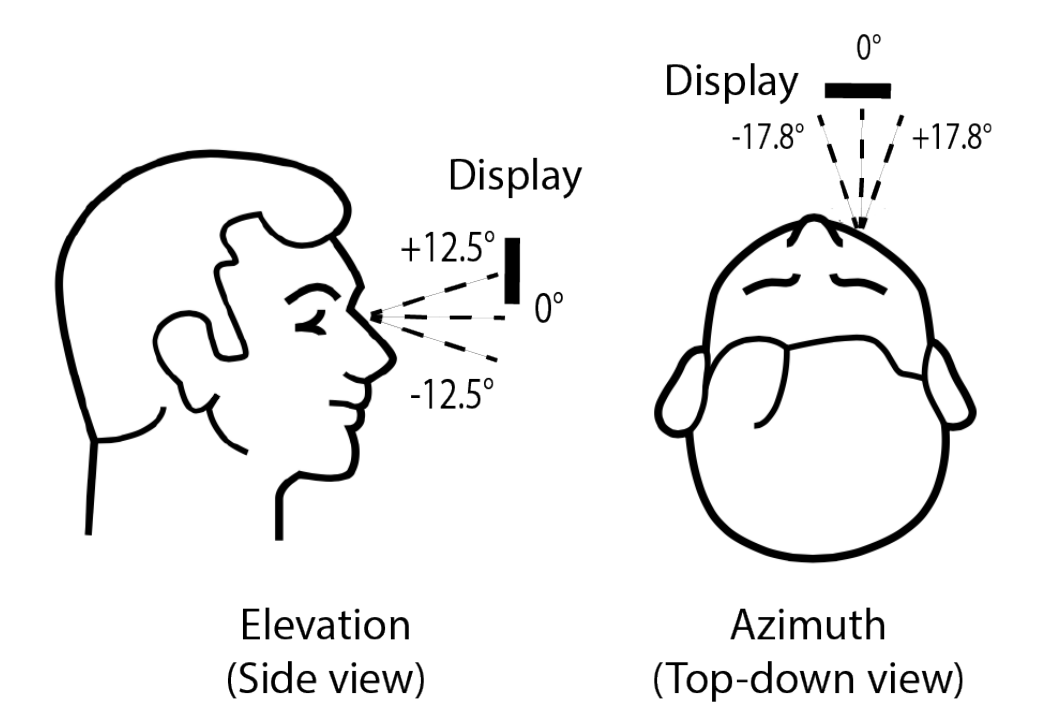}
    \caption{Illustration of three elevation and azimuth
angles relative to the right eye (by Chua et al. with permission) \cite{chua}.}
    \label{fig:angles}
\end{figure}

Typically, for defining the positioning of a display for head-stabilized       \cite      {billinghurst1998spatial} HWDs, the positions are defined in angles relative to the head. We will use this terminology throughout the paper. There are two primary angles: azimuth (horizontal) and elevation (vertical) (Figure \ref{fig:angles}), where 0° azimuth and 0° elevation represent the center in front of the user, also called as PPOG (Principal Position Of Gaze) .

\subsubsection{Azimuth (Horizontal Angle)}
Azimuth refers to the horizontal angle relative to the user’s center of gaze. This angle is measured from the user’s direct forward view (0°) and extends outward in both directions: positive azimuth values denote positions to the right, and negative values denote positions to the left. Azimuth determines how far left or right an object appears relative to the wearer’s gaze.

\subsubsection{Elevation (Vertical Angle)}
Elevation defines the vertical positioning relative to the user’s forward view. This angle is also measured from the central gaze (0°). Positive elevation values indicate positions above the gaze, while negative values represent positions below it. Elevation dictates how high or low an element is within the wearer’s visual field.

\subsection{Defining Standard Display Positions}
\begin{figure}[ht]
    \centering
    \includegraphics[width=0.6\linewidth]{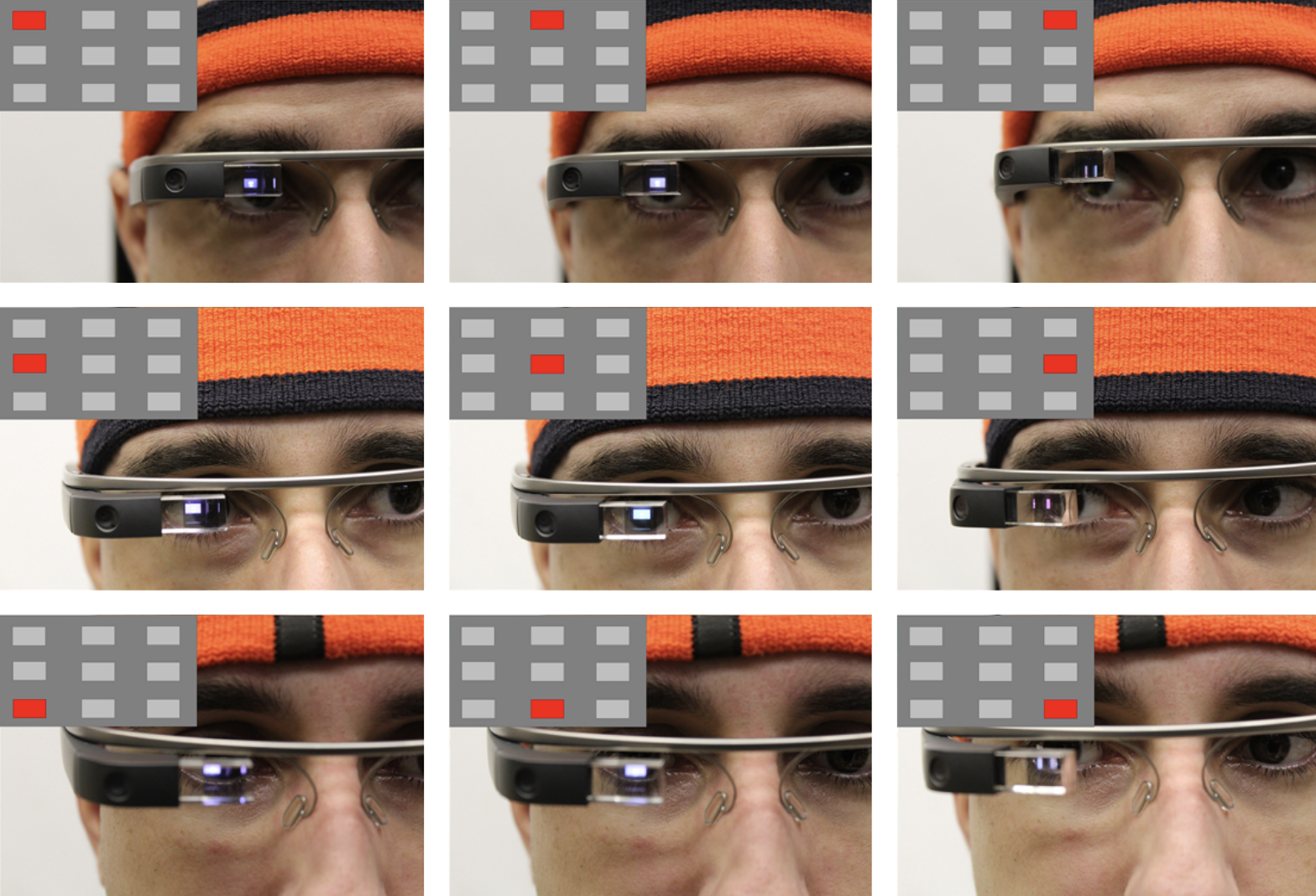}
    \caption{Nine display positions on a
monocular OST-HMD studied by Chua et al. (reproduced with permission) \cite{chua}. The red rectangle on the diagrams located at the top left corner of each image indicates the
display position from the user's point of view.}
    \label{fig:chuaPositions}
\end{figure}

By combining azimuth and elevation, we can define nine standard positions within the wearer’s field of view. 
These positions represent the most common placements for visual elements in a HUD.
These terms were used in Chua et al.'s study (see Figure \ref{fig:chuaPositions}), and we adopt them here for convenience:

\begin{itemize}
    \item \textbf{Center-Center}: Directly in front of the user’s gaze (0° azimuth, 0° elevation).  
    \item \textbf{Top-Center}: Directly above the user’s gaze (0° azimuth, +elevation).  
    \item \textbf{Bottom-Center}: Directly below the user’s gaze (0° azimuth, -elevation).  
    \item \textbf{Center-Left}: To the left of the user’s gaze (-azimuth, 0° elevation).  
    \item \textbf{Center-Right}: To the right of the user’s gaze (+azimuth, 0° elevation).  
    \item \textbf{Top-Left}: Above and to the left (-azimuth, +elevation).  
    \item \textbf{Top-Right}: Above and to the right (+azimuth, +elevation).  
    \item \textbf{Bottom-Left}: Below and to the left (-azimuth, -elevation).  
    \item \textbf{Bottom-Right}: Below and to the right (+azimuth, -elevation).
\end{itemize}

\subsection{Task Types}
Optimal positioning also heavily depends on the specific task or use case for which the HWD is designed. Devices intended for everyday wear must accommodate a wide range of tasks. For consistency, HWD use cases can be categorized into four groups       \cite      {lin_towards_2021}:
\begin{itemize}
 \item \textbf{Single tasks}: Activities where the user's complete attention is on the HWD, such as reading a book, web browsing, or watching a video.

\item \textbf{Background tasks}: Situations where the user's primary focus is on the physical world, but they intermittently check the HWD for notifications, such as the presence of incoming calls or texts.

\item \textbf{Alternating tasks}: Scenarios where users frequently switch focus between the virtual display and physical reality, such as order picking (e.g., locating items in a warehouse) or task guidance (e.g., assembling a car engine). In this case, the tasks performed in the real world and on the HWD are directly related.

\item \textbf{Dual tasks}: Cases where the HWD is integrated into the user’s physical-world activity. Unlike alternating tasks, the tasks on the HWD and in the physical world are unrelated and require a shift in mental focus, such as responding to and acting on notifications (not just noting their presence, as in background tasks) on the HWD while driving.

\end{itemize}

%thad: updated in plane 4/22 9am

Note that the above classification of tasks has the implicit
assumption that the content provided by the HWD is distinct and
distinguishable from the content of the physical world. Often
proponents of augmented reality suggest that the ideal AR display is
one where the real and virtual are indistinguishable. In such a
situation the virtual and physical tasks above could be merged into
one task. We do not include such situations in our task descriptions
as this ``ideal'' situation is not possible with a monocular HWD as
the non-display eye would, by definition, see the unadorned physical
world. (Even binocular optical see through displays are unlikely to
completely meet this ideal. Consider the situation where a virtual
object must occlude a physical object. There is currently no practical
way for the HWD graphics to completely obscure objects in the physical
world while still matching the lighting effects in the physical
world.)

\section{Choosing a field of view for everyday use}

The horizontal field of view (FOV) is a critical design consideration when designing and positioning displays. The size of the display can impact performance at different positions. To measure the FOV of any device, not just HWDs, the following formula is typically used:

\[
V = 2 \cdot \tan^{-1}\left(\frac{S}{2D}\right)
\]

where \( D \) is the viewing distance and \( S \) is the horizontal
screen width in centimeters       \cite      {kaiser2004joy}. For
tasks like reading text on mobile devices, the typical viewing
distance is around arm's length, approximately 36.2 cm (14.25'') \cite
{bababekova2011font}. However, laptops are typically viewed at 18''
(45.7 cm). We can use these approximations to explore the
FOVs of common situations in everyday life.

Depending on the distance in which it is held, a typical newspaper
column of text (width 4.6 cm \cite {wikipedia_column_inch}) has an FOV
of 5.76° - 7.27°, which we set as our minimum horizontal FOV. The
iPhone 6, held in portrait orientation with a 5.8 cm wide display
\cite {iPhone_6}, provides an FOV of 9.2°, while the iPhone 16 Pro Max
(display width 7.32 cm \cite {iPhone_16_pro_max}) offers an FOV of
11.55°. The video iPod (4th Gen Nano), held in landscape mode with a
horizontal screen width of 9.07 cm \cite {video_iPod}, has an FOV of
14.28°.

Tablets, presumably, make a different trade-off between size and
portability. The iPad Mini 2 (2017 model, width 13.47 cm \cite
{iPad_Mini_2}) achieves an FOV of 16.76° - 21.07°, while the iPad Air
2 (2017 model, width 16.95 cm \cite {iPad_Air_2}) results in an FOV of
21.01° - 26.35°. The latest iPad Pro with the M4 (13") model (width
21.55 cm \cite {iPad_Pro_M4}) provides an FOV of 26.53 - 33.15°. In
comparison, a 12.5'' laptop screen (27.9 cm horizontal), viewed at the
typical 18'' (45.7 cm) distance results in an FOV of 34°.

While less portable devices seem to find acceptance at 15° - 35°
horizontal field of view, smartphones are the dominant computer user
interface currently. This dominance suggest that having a horizontal
FOV of less than 15° is an acceptable compromise given the other
benefits that these mobile devices provide.

For home theater-style entertainment, the Society of Motion Picture and
Television Engineers (SMPTE) recommendations suggest a 30° FOV
for optimal viewing \cite{smpte}. THX \cite{thx}, however, recommends
35° for movie theaters. These recommendations align with the maximum
HWD FOV suggested by ophthalmologists to prevent strain
\cite{davis1997visual,Peli_1999} and align with the typical eye range
of motion before requiring head movement
\cite{sidenmark2023coordinated}. Indeed, such recommendations match
everyday observations in movie theaters. Most patrons sit in the
middle of the theater (approximately 35° as measured by the last
author) as opposed to the front row.  In contrast, IMAX theaters
emphasize their large FOV, which ranges from a minimum recommendation
of 55° to a maximum of 115°.

Google Glass XE, was introduced with a horizontal FOV of 12°, which proved
sufficient for microinteractions during all-day wear. The latest Vuzix
Z100 \cite {vuzix_ultralite} ships with a 24° FOV. As waveguide
technology improves, optics manufacturers like Dispelix, Digilens, and
Lumus \cite {Dispelix, Digilens, Lumus} now offer a wide range of
FOVs, with some exceeding 50°. Such relatively large FOVs may be more
suited for immersive binocular devices.  However, these devices,
typified by Meta Orion \cite{MetaOrion}, are constrained by thermal,
power, and weight issues and are not yet ready for all day wear.

While a higher FOV enhances immersion and improves the user experience
for tasks such as gaming, spatial modeling or multitasking with
multiple virtual interfaces, it is often impractical for devices
designed for all-day wear, at least at the time of writing. Current
technology struggles to balance high FOV with contraints such as bulk,
weight, and power consumption, which can compromise comfort and
battery life over extended use.

\subsection{Eye movement}
When using a desktop display, both head and eye movement can be used to attend different positions of the screen.  However, an HWD's image is fixed relative to the head such that the eyes alone must move to see the content of the screen. Moving the eyes to extreme
angles can significantly affect visual comfort. Although the physical
range of eye rotation spans up to $\pm 50^\circ$, studies indicate
that the eyes rarely rotate beyond $\pm 30^\circ$ relative to the head
\cite{land2009looking}. The range of everyday eye movements would
suggest further limiting the field to about $30^\circ$ (e.g., $\pm
15^\circ$) \cite{davis1997visual, peli_visual_1998}. The eyes tend to
stay within a comfortable range and often wait for the head to re-align
for extended shifts. Hence, gaze shifts within 20° are predominantly
achieved through eye movement, with the eyes contributing over 90\% of
the shift for targets within a $25^\circ$ range
\cite{freedman2008coordination, goossens1997human,
  land2009looking}. For gaze shifts within the $25^\circ$--$50^\circ$
range, the eyes and head coordinate to achieve the target
comfortably. When shifts exceed $50^\circ$, additional body movements
such as torso rotation become necessary for effective and comfortable
viewing \cite {morimoto2005eye}. Research by Sidenmark et al.  \cite
{sidenmark2023coordinated} emphasizes that these dynamics form a
baseline for evaluating whether head and eye coordination in virtual
environments mirrors natural settings. These findings delineate three
distinct gaze shift strategies:

\begin{enumerate}
    \item \textbf{Eyes-only range ($<25^\circ$)}: Eye movement dominates.
    \item \textbf{Eyes-and-head range ($25^\circ$--$50^\circ$)}: Both eyes and head contribute.
    \item \textbf{Eyes-head-torso range ($>50^\circ$)}: Full-body coordination is involved.
\end{enumerate}

\begin{figure}[ht]
    \centering
    \includegraphics[width=0.5\linewidth]{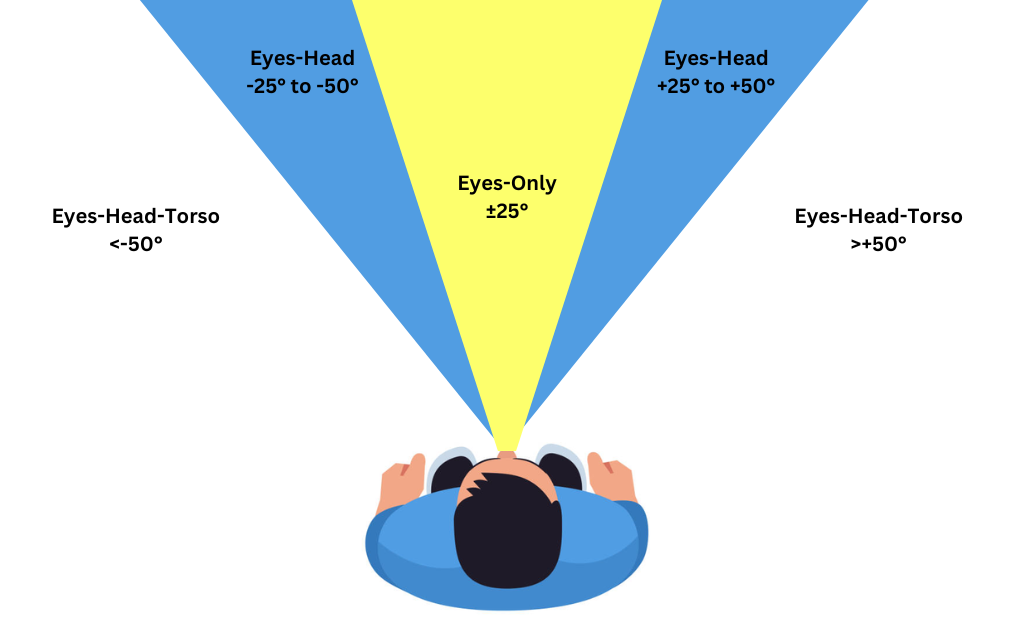}
    \caption{Illustration of eye ranges defined by Sidenmark et al. \cite{sidenmark2023coordinated}.}
    \label{fig:eyeRanges}
\end{figure}

These results suggest that, if the user is expected to be attending
the physical world at the PPOG, then virtual content aligned with
respect to the reference frame of the head should ideally be
positioned within an arc of \(\pm 25^\circ\) (a total of \(50^\circ\))
from the central visual field such that the user can glance at it
without engaging the natural tendency to move the head to view the
content. However, if the user is mainly attending virtual content (as
with single tasks such as reading a book), then the content should be
limited to $<25^\circ$.

%This indicates that there is a maximum range within which the eyes can move relative to the head comfortably, and this must be considered when determining how far displays should be positioned. 

%Optical see-through devices have already been shown to increase visual load and decrease awareness       \cite      {huckauf2010perceptual, orlosky2014managing}, which can, in turn, affect user comfort.  

Note that large FOV displays almost require that the active virtual
image region overlaps the principal position of gaze
(PPOG). Sustaining the gaze to the side 40° or more is painful for
most users. Thus, a 50° FOV display offset to the side so as not to
intersect the PPOG would be placing a signficant amount of content in
this painful region. We must assume, then, that large FOV displays
(>40°) will overlap the PPOG, which leads to another potential
issue. Ongoing studies suggest the user may be annoyed with the sudden
display of graphics overlapping the PPOG (see below).

While studying FOV for captioning using HWDs, Britain et al.
\cite {david} reported that a field of view (FOV) between 20° and 30°
is preferred for captioning for deaf and hard-of-hearing users
during group conversations. This study assumed a HWD with a virtual
image centered on the PPOG. However, an ongoing study building on Britain’s
work by Kwok et al. suggests that a 15° FOV image may be ideal for
captioning at a 9° offset towards the ear. This 9° offset towards the
ear was chosen based on the findings of Arora
\cite{arora2024comfortably} which found that a monocular display on
the right eye between -20.2° and +8.7° can be annoying due to the
edges and distortions caused by optical combiners. Kwok found that, for
larger FOVs, users often move their eyes or head frequently to read
captions across the wide field, causing strain over extended
use. However, at a 10° FOV, users had difficulty reading the captions
as they constantly had to move their eyes from line to line as there
was a limited amount of content per line. In addition, participants
used the captions to provide context if they lost track of the
conversation. At 10° FOV the display provided only a few words of
context making it harder to follow the conversation.  

{\bf Thus, if a goal is to avoid the PPOG for devices designed for all-day
wear while still providing a good experience for reading text (such as
captioning for wearers who are deaf or hard of hearing), a 15° FOV is
recommended. If avoiding the PPOG is not an issue, then a 25° FOV is
recommended. }

These results align with the horizontal FOVs of common devices
mentioned earlier.  Tablets, which are viewed centered around the PPOG
presumably, typically have a FOV between 20° and 30°, similar to the
preferences reported by Britain et al. \cite {david}. However,
pocketable devices such as smartphones and video players, must balance
portability and FOV and typically range from 9° to 14° horizontal
FOV. While HWDs must balance different trade-offs, the current
compromise with smartphones, made by a mature industry in one of the most popular
consumer electronics segments, is reassuring in that it corresponds
with the 15° recommendation above.

%(and maybe add 50°/\(\pm 25^\circ\)  FOV/FOR from Sidenmark?, or just a better conclusion?)

\section{Considerations for display positioning}

There are several factors to consider when positioning the display. It
should be placed centrally enough for the user to comfortably and
quickly view the content, yet unobtrusively enough to avoid constant
interruption. Research has been conducted to evaluate specific
scenarios and tasks, assessing different considerations for display
position. We have categorized these considerations into four broad
areas: interruptions, comfort, performance, and social perception.

\subsection{Interruptions}

Head-worn displays can be interruptive, especially when they display
content without being explicitly triggered by the wearer (such as with
notifications or proactive AI). Extensive research has been conducted
on the use of these displays in specific scenarios where minimizing
interruption is important. These include head-up displays (HUDs) in
the aviation industry for piloting planes, stationary HUDs and mobile
head worn displays (HWDs) while driving, and mobile HWDs while
walking.

\textbf{Aviation} \\
In aviation, researchers have studied the phenomenon of cognitive
capture (also known as cognitive tunneling), a challenge particularly
prevalent in early aircraft HUDs. Cognitive capture refers to the
difficulty users experience when shifting attention from the HUD to
other elements in the visual environment. Unlike typical attention
shifts, such as moving between an instrument panel and the external
environment, cognitive capture occurs entirely within the same field
of view and does not require changes in fixation, visual
accommodation, or convergence. Fischer et al.
\cite{fischer1980cognitive} and Weintraub et
al. \cite{weintraub1985head} identified that center-aligned HUDs,
although effective, were more prone to cognitive capture, impairing
pilots' ability to maintain control. Foyle et al. \cite{Foyle} found
that aligning HUDs at the PPOG exacerbated cognitive
capture. Participants in flight simulations struggled to maintain
altitude when HUDs were center‑aligned compared to when they were
%thad: display edges - was the center of the display offset 8.14 degrees?  Or was the corner closest to PPOG 8.14 degrees?  What was the display FOV? 
offset diagonally by 8.14° or 16.28° above and to the left of the
PPOG. Dowell \cite {dowell} further demonstrated that HUDs within an
%had: display edges - is 8 degrees the pixel closest to PPOG?  If so we need to say so.
8° radius of the primary point of interest caused cognitive capture,
while those positioned beyond this radius reduced such effects.  Note
that these studies, and the older automotive studies below, did not
report FOVs of the displays as they focused on presenting small
symbols or a few characters of text to the pilot as opposed to a
rectangular raster display. 

\textbf{Automotive} \\ In automotive HUDs, situational awareness has
been a focal point of research. Gish and Staplin \cite {gish1995human}
studied the positioning of HUDs displaying warning signs and concluded
that HUDs performed best when placed near the driver’s line of sight,
with optimal offset for the edge closest to PPOG %thad: display edges 
being between 5° and 10°. 
Subsequent studies by Yoo
\cite{yoo1999display}, Watanabe \cite{watanabe1999effect}, and
colleagues, who examined 15 different positions, corroborated this
finding, revealing that HUDs with their edge %thad: display edges
positioned 5° to the right of center yielded the best performance. Tsimhoni et al.  \cite
{tsimhoni2000display, tsimhoni2001detecting} extended this research by
using text in the HUD rather than graphical symbols. Tsimhoni showed
that while reading response time increased with horizontal
eccentricity, detection time remained unaffected. Participants
preferred the three center positions in the middle row, with the most
preferred position being 5° to the right of center (in this discussion,
center refers to the PPOG of users' right eye). Further studies
reinforced these conclusions \cite{cars_chao2009see, cars_fukano,
  cars_inuzuka, cars_Lino, cars_park}. These findings have shaped
industry recommendations, such as the Society of Automotive Engineers’
(SAE) guidance to place critical warnings within 10° of the driver’s
line of sight \cite {mcgehee2002human} and the
%thad: display edges - I assume that 10 degrees is the closest pixel?
%Answer: Yes, but I dont think we should write "Closet pixel"specifically, i think its pretty clear and itll just add more clutter. Can still add it you think its unclear for this one
Japanese Automobile
Manufacturers Association’s (JAMA) advice to ensure forward visibility
when viewing displays \cite {nakamura2008jama}. Recent studies have
studied see-through heads-up displays in the windshield of automobiles
compared to traditional Heads Down displays (HDDs). Several found that
drivers performance decreased as the HUD moved far away and drivers
% thad: does far away mean far from PPOG? or rendered in the distance?
preferred HUD imagery that was closer to them \cite
{topliss2019evaluating, horrey2003does, jose2016comparative}. Kim et
al. studied the visual annoyance potential of see-through HUDs by
investigating driver gaze distribution across different types of
objects on the road. They discovered that such interfaces can be
informative or annoying depending on the perceptual forms of
graphical elements presented on the displays. See-through head-up
displays (HUDs) can direct the driver’s attention away from critical
road elements that are not augmented by the displays \cite
{kim2022assessing, kim2018quantifying}. Chua et al.  \cite {chua}
studied mobile head worn displays using Google Glass by
investigating nine different display positions (Figure
\ref{fig:chuaPositions}) and their effects on performance and
usability in a dual-task scenario.

Chua investigated positioning the center of display at 3 elevation (+12.5°, 0°, -12.5°) and 3
azimuth angles (-17.28°, 0°, +17.28°) to generate the nine positions (see Figure \ref{fig:linPositions}).
Participants responded to three
types of notifications displayed on the HMD while performing a
visually intensive single task of driving in a simulator. Despite the
center-center position being the fastest for reaction time, it was not
the most comfortable or preferred. Some participants found the
center-center and all bottom positions not comfortable for the primary
task. While notifications at the center-center and bottom-center
positions were noticed more quickly, the top-center and center-right positions
were more comfortable and unobtrusive. Specifically, the center-right
position provided the best balance between performance and
usability. Chua's findings were consistent with those of Yoo,
Watanabe, and Tsimhoni \cite {yoo1999display, watanabe1999effect,
  tsimhoni2000display, tsimhoni2001detecting}, who all found that
users preferred HWDs offset slightly to the right of center for
comfort.

\begin{figure}[ht]
    \centering
    \includegraphics[width=0.8\linewidth]{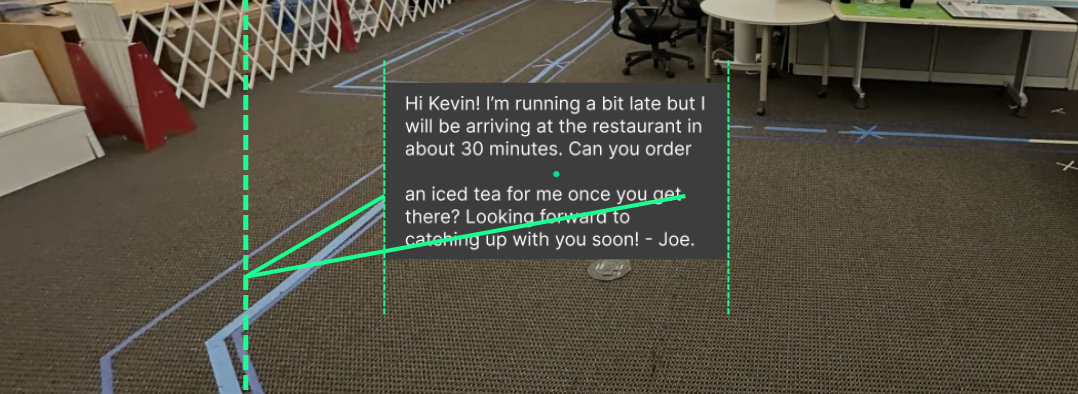}
    \caption{Simulated text message in HWD offset from PPOG while user walks on pre-defined path.}
    \label{fig:mosur}
\end{figure}

\textbf{Walking} \\ Oulasvirta et al. \cite{oulasvirta2005interaction}
studied conversing, walking and riding transportation while
interacting with mobile devices. These tasks compete for cognitive
resources, particularly attention, and the research found that users
typically engage with the mobile device in four second bursts while
walking, due to the constant switching of attention between mobile
tasks and environmental distractions. Previous research on mobile
devices shows that walking affects reading \cite
{schildbach2010investigating, vadas2006reading}. However, it remains
unclear if this effect holds for reading on HWDs. As OST-HWDs do not
fully occlude the user’s field of view, they hold promise for reading
while walking. Related work indicates that smart glasses enable higher
situational awareness than smartphones \cite
{orlosky2014managing}. Lucero and Vetek \cite {lucero2014notifeye}
investigated background tasks of getting social network notifications
on a OST HWD while walking on a busy street. They found that receiving
minimalist notifications on smart glasses while walking did not
interfere with the user’s vision. More recently, Mosur et al.  \cite
{mosur2024stepping} studied interruptions by asking users to read
notification text on a Meta Quest 3 \cite {metaquest3} using
color video pass through to simulate a 15° FOV augmented reality display
while walking (Figure \ref{fig:mosur}). The study simulated hardware
failures, such as the simulated 15° FOV screen suddenly flashing
white. Participants were able to walk and read simultaneously with
minimal missteps, showing no significant impact from the 15° FOV
field-of-view HWD, even when emulating a malfunction. As suggested by
Oulasvirta et al.'s study on smartphones, participants seem able to
transition from the real and virtual content quickly enough to have
minimal effect on their walking.

%Ongoing studies and pilots based on Mosur's work suggest that larger
%FOVs show a similarly minimal effect in a following scenario, though
%it remains unclear how larger FOVs influence walking behavior
%real-world scenario with many potential hazards.

Building on the work of Mosur et al. \cite{mosur2024stepping}, an
ongoing study by Kimmel et al. focuses on larger FOV AR displays.  Participants follow
a moving virtual avatar with constantly changing speed in an office
environment while wearing a Meta Quest 3 \cite{metaquest3} mixed
reality headset with video pass through to simulate different types of
HWDs. The study simulates worst-case visuals by displaying opaque
horizontal and vertical bands of varying thicknesses (as well as large
squares) while the user is walking. To be specific, a 60° horizontal
band would extend across the full extent of the Quest 3's FOV but is
centered such that 30° appears above the PPOG and 30° appears
below. The wearer can ``look around'' above and below the
band. Although the study is still ongoing, initial pilot tests have
shown that participants can navigate without difficulty with occluding
bands up to 60° of width. Additionally, horizontal banding was
strongly preferred over vertical banding, as vertical eye and head
movement was noticeably more comfortable than horizontal eye movement
to ``look around'' the occlusion. In other words, while sudden visuals
appearing in the PPOG may be annoying to users, these pilot results
suggest they can still navigate while walking as long as the centered
vertical FOV is less than 60° such that they can see under the
graphics with a quick movement of the head and eyes.
%Compare these results to studies testing the use of head-up displays
%while driving a car or piloting a plane (see below).

In summary, aviation studies suggest staying outside an 8° radius from PPOG. Automobile studies also support an offset. 
Chua's driving simulation study using modified versions of Google Glass suggests that a PPOG-aligned HWD can be interruptive and suggests an offset to the right, aligning with previous studies. Interestingly, Google Glass's normal top-center image location was such that all pixels were more than 8° away from the PPOG. Given that thousands of people have worn Glass as part of their everyday lives or jobs between 2012-2023, it provides some evidence that 8° is sufficient. Combining these results {\bf from the perspective of interruption, the virtual image of a HWD should be positioned outside an 8° radius from the PPOG and, based on Chua's studies, potentially in the center-right position.}

% \begin{figure}[ht]
%     \centering
%     \includegraphics[width=0.6\linewidth]{images/notifEye.png}
%     \caption{The NotifEye UI shown on their glasses’ display.       \cite      {lucero2014notifeye}}
%     \label{fig:notifEye}
% \end{figure}

% \begin{figure}[ht]
%     \centering
%     \includegraphics[width=0.4\linewidth]{images/aircraftHud.png}
%     \caption{Center-aligned HUD in aircraft from Fischer et al       \cite      {fischer1980cognitive}}
%     \label{fig:planeHud}
% \end{figure}

% \begin{figure}[ht]
%     \centering
%     \includegraphics[width=0.4\linewidth]{images/car_hud.png}
%     \caption{Locations studied by Tsimhoni et al       \cite      {tsimhoni2001detecting}}
%     \label{fig:outerBounds}
% \end{figure}

% \subsection{Comfort \& Performance}
\subsection{Comfort}

If an HWD is designed to be worn like normal eyeglasses in everyday life, the display may often be off to preserve battery life.  Thus, for the purposes of discussing comfort, we must consider both the situation when the display is on and when it is off. Comfort with the display turned on focuses on the user's experience while consuming content on the virtual display, whereas comfort with the display turned off considers the potential visual disturbances caused by the optical combiners, which may appear as a blurry smudge on the glasses or rainbows caused by stray light interacting with the optics. 

\subsubsection{Active Display}

Several studies have focused on user preferences for the position of
display. In the previously mentioned study done by Chua et
al. \cite{chua}, the center-right position was rated the most
comfortable by users, followed by top-center. Differences in comfort
scores reached statistical significance when comparing center-right to
bottom-left and top-left.  Interestingly, user preference ratings also
placed center-right first, followed by top-center.  There was a
statistically significant difference between center-right and
bottom-left and top-left. While center-center had the fastest reaction
times, it rated fourth (of nine) in comfort and preference. The next
fastest position, bottom-center, was ranked fifth in comfort and
preference. Positions on the left trended to be rated lower than all
other positions.

%thad: display edges
% Answer - added details for chua above when chua was mentioned first
% Answer - for zhen, mentioned position of glass
Zheng et al.       \cite      {dual_Zheng} also studied HWD placement in guided automotive maintenance scenarios, comparing above-line-of-sight (e.g., Google Glass) and in-line-of-sight (e.g., Epson Moverio) positions. Note that Google Glass has a display positioned such that the bottom-most edge is about 8° above the line of sight.  Similar to Chua, Zheng found the in-line-of-sight position to have higher performance; however, twice as many users preferred the above-line-of-sight position over the in-line-of-sight position. 

\begin{figure}[ht]
    \centering
    \includegraphics[width=0.8\linewidth]{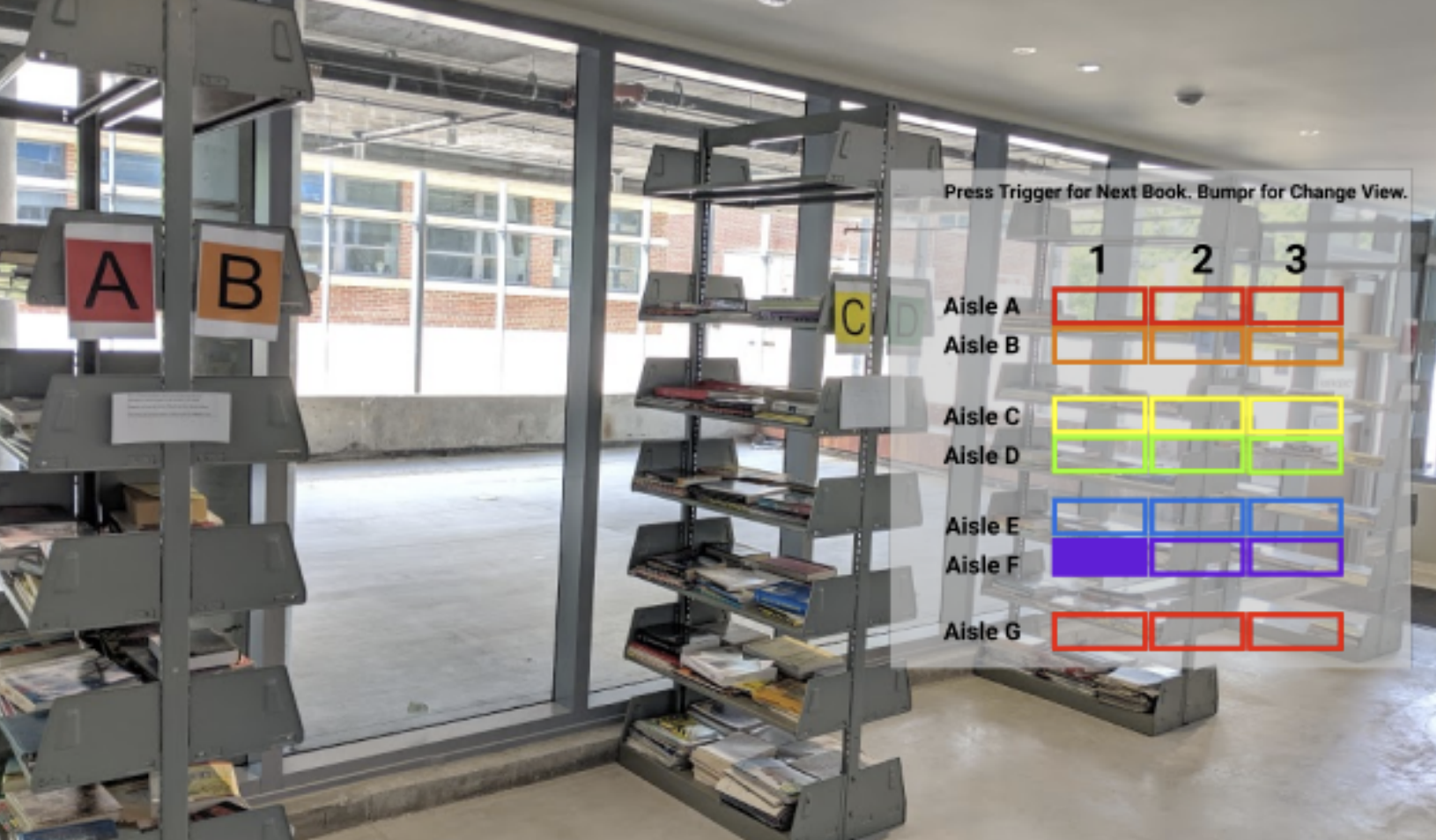}
    \caption{Simulated examples of participant's view in Lin et al.'s task (with permission) \cite{lin_towards_2021}.}
    \label{fig:linTasks}
\end{figure}

\begin{figure}[ht]
    \centering
    \includegraphics[width=0.8\linewidth]{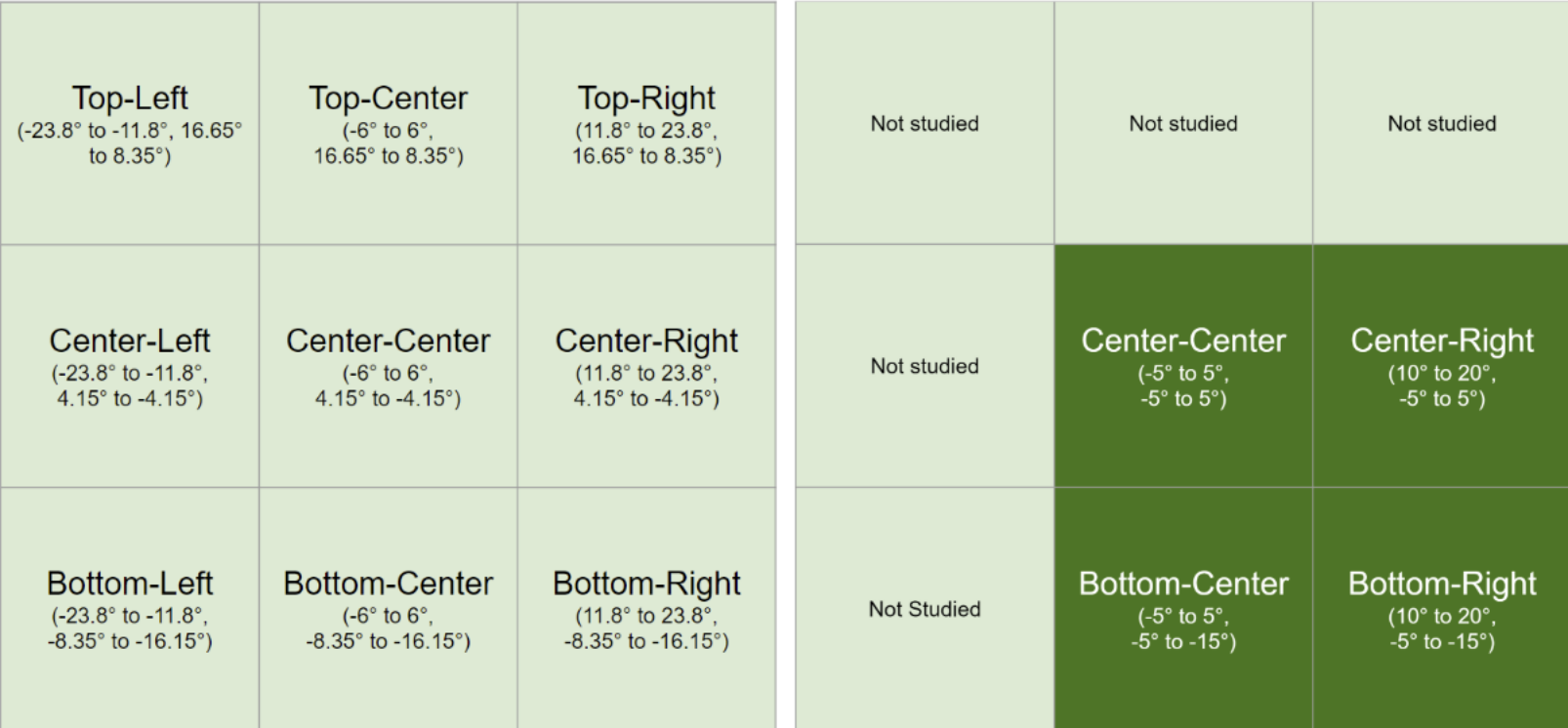}
    \caption{Left: Nine HWD positions tested by Chua et al. \cite{chua}. Right: Display positions tested by Lin et al. \cite{lin_towards_2021}.}
    \label{fig:linPositions}
\end{figure}

Lin et al. selected high-performing positions from Chua's study and examined their effect on walking tasks. Participants performed an order-picking task in simulated warehouse conditions (Figure \ref{fig:linTasks}) using the Magic Leap 1       \cite      {MagicLeap}. Lin tested four interface positions in a within-subjects study (see Figure \ref{fig:linPositions}): center-center (-5° to 5° horizontal, -5° to 5° vertical), center-right (10° to 20° horizontal, -5° to 5° vertical), bottom-center (-5° to 5° horizontal, -15° to -5° vertical), and bottom-right (10° to 20° horizontal, -15° to -5° vertical). Lin found comfort for all positions was similar with the exception of the bottom-right, which was ranked significantly lower than center-center.

To further investigate the right offset position suggested by Chua, Lin et al.       \cite      {lin_towards_2021} studied the comfort of HWD when the display was offset to the right. Lin studied different display positions with a field of view (FOV) of 9.2° horizontally, centered at offsets of 0°, 10°, 20°, and 30° (i.e., -4.5° to 4.6°, 5.4° to 14.6°, 15.4° to 24.6°, and 25.5° to 34.6°). 
Lin used a MicroOptical SV6 opaque monocular    \cite      {MicroOpticalSV6} HWD controlled by a Raspberry Pi for a button-pushing experiment conducted while standing. The experiment required participants to identify which button to press on a 3x4 shelving array using the HWD. Results showed that the 10° offset was most preferred, with 0°, 10°, and 20° being more comfortable than 30°. Additionally, 0° and 10° caused less eyestrain than 20° and 30° and were more visually comfortable than 20°. There were no significant differences in workload, speed, or accuracy across conditions.

Building on Chua and Lin's       \cite      {chua} work, Haynes et al.       \cite      {haynes2018effects} conducted a study to investigate the right offset position simulating  long-form text reading on HWDs. Haynes chose reading as a suitable single task for this evaluation since it demands more attention than watching an image or video. Using an emulated 9.2 x 16.3 degree FOV display for six 30-minute reading sessions, participants rated comfort when the display was centered at four horizontal positions: 0°, 10°, 20°, and 30°. The intent was to test the feasibility of the right offset under the most demanding conditions compared to the center position. Results indicated that 0°, 10°, and 20° offsets were rated as more comfortable, with 0° and 10° being slightly more favorable than 20°. Post hoc analysis confirmed that offsets of 0°, 10°, and 20° caused less eyestrain than 30°. Differences in workload and objective performance measures (reading speed, accuracy, and comprehension) across conditions were minimal. Interpreting the results for this discussion, while the right offset was comfortable for users, display pixels should be restricted from being more than 
% 24.6° 
30°
toward the ear (right) for tasks lasting 30 minutes or more. 

\begin{figure}[ht]
    \centering
    \includegraphics[width=0.4\linewidth]{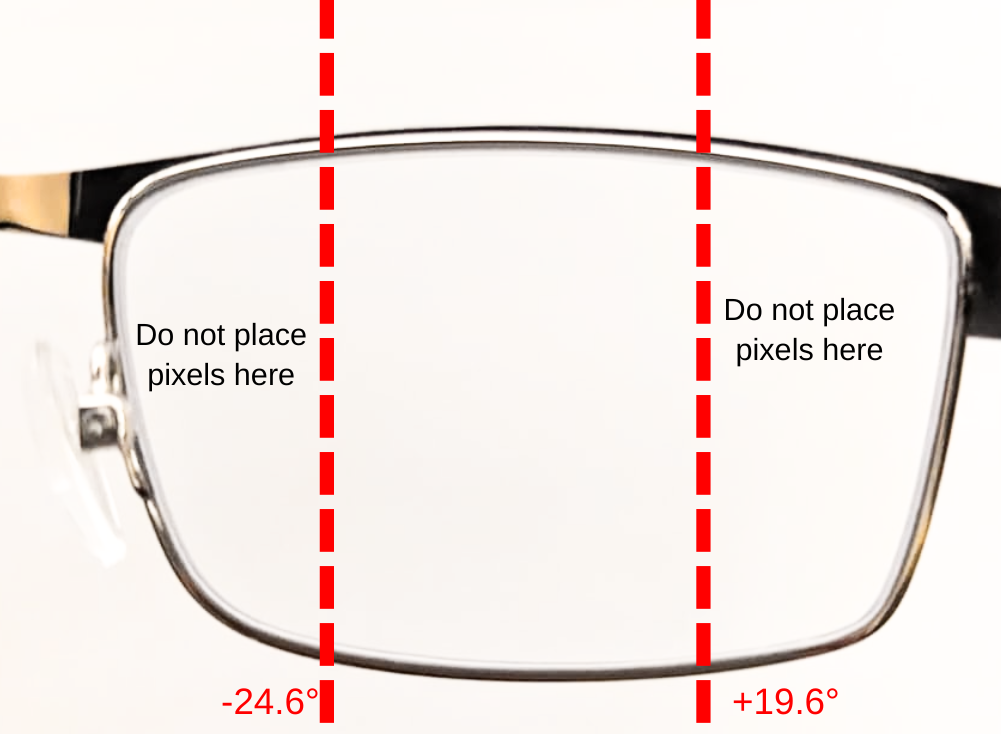}
    \caption{Outer bounds to display pixels.}
    \label{fig:outerBounds}
\end{figure}

Song and Arora et al.       \cite      {yukunParthSherlock} conducted two additional studies building on this methodology. Similar to Haynes, they used reading as a suitable single task and replicated the procedure. The first study offset the display to the left, and in the second study, they compared both directions with slight offsets to normalize the data with the center position and combined the results. Their findings suggest that reading on right-eyed displays with pixels between -24.6° (towards the nose) and +19.6° (towards the ear) is comfortable for most users (see Figure \ref{fig:outerBounds}). 
In their analysis, Song and Arora et al. showed that discomfort increased quickly further than 24.6° toward the ear, suggesting that 24.6° itself was still satisfactory for the average user.  

% However, both 
Both this study and Haynes's showed that some individual participants had a marked increase in discomfort at 24.6°. Thus, we adopt 19.6° as a more conservative boundary. Interestingly, a display offset 30° toward the ear was considered barely acceptable.

Mosur et al.       \cite      {mosur2024stepping} further studied this range by simulating text messages on video pass though HWD Meta Quest 3       \cite      {metaquest3}, and  offsets varied to move away from the PPOG, but within the suggested outer bound range of -24.6° and +19.6°, plus some best and worst cases for comparison and found that there was no significant difference between difference offsets when compared to central 0 position. This aligns with Sidenmark's suggested gaze shifts of $\pm$ 25° for eyes-only movement       \cite      {sidenmark2023coordinated} and film viewing guidelines  mentioned above. 

{\bf Hence, when considering comfort while using an active display
  mounted on the right eye, it is recommended to position the display
  between -24.6° and +19.6° for extended use and up to 30° toward the
  ear for shorter interactions for comfort when the display is on. User
  preference tends to follow comfort ratings in the literature. Both
  favor the center-right position repeatedly. If we combine these results
  with the previous section on interruption, we begin to see a pattern
  supporting a vertically centered image offset by 8° towards the
  ear (i.e., no pixels nearer the PPOG than 8°). We also begin to see a trend
  against positioning the image in the diagonals (top-left, top-right,
  bottom-left, bottom-right). }

\subsubsection{Inactive Display}
\begin{figure}[ht]
    \centering
    \includegraphics[width=0.7\linewidth]{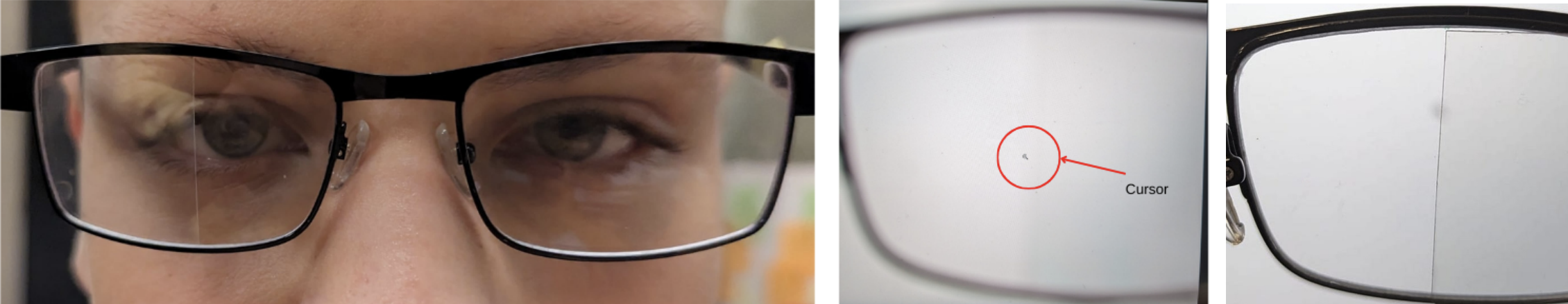}
    \caption{Emulated optical combiner. Left: The emulated combiner is placed in front of the participant’s right eye. Middle: First person view through the emulated combiner focusing on a bright white screen. Right: First person view focusing on the combiner (note that a wearer can not focus close enough to see the combiner's edge clearly).}
    \label{fig:combiner}
\end{figure}

\begin{figure}[ht]
    \centering
    \includegraphics[width=0.7\linewidth]{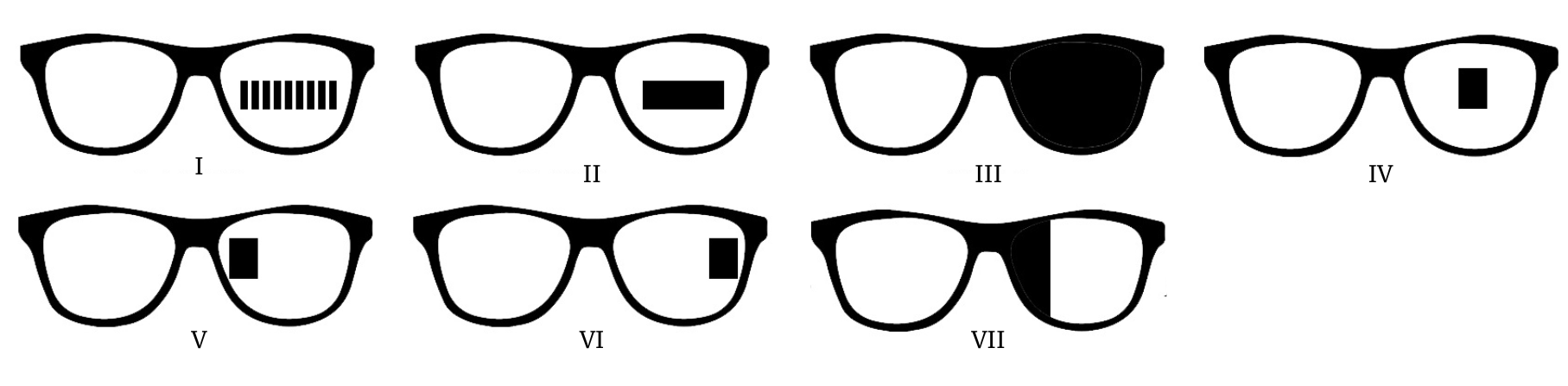}
    \caption{Configurations of optical combiners considered.}
    \label{fig:shapes}
\end{figure}

HWDs can be visually uncomfortable when the display is turned off as
optical combiners can appear as a blurry smudge or grey area on the
glasses. To preserve power during everyday use, HWDs may be turned off most of
the time (such as with Google Glass or the Vuzix Z100).  Therefore, it
is necessary to test how the appearance of the unlit optical combiner
might affect positioning of the virtual image (note we do not consider
the effect of stray light on the combiner, such as the ``rainbows''
that can occur with diffraction designs). Song and Arora evaluated
different shapes of optical combiners at various positions
\cite{artGallaryFirst}. The shapes and offsets under consideration
were based on those commonly used in HWDs for everyday wear at that
time (see Figure \ref{fig:shapes}). The glasses used semi-transparent
films to emulate optical combiners in seven different locations
(Figure \ref{fig:combiner}), and the study asked users to rank these
models by preference. Models VI and VII were the most preferred, with
model VII slightly outperforming VI. Song and Arora suggest that
this preference was likely due to model VII having the fewest
edges. It is worth noting that both the highly ranked models had a
lateral offset, which aligns with the preferences reported by Chua et
al. \cite{chua} and others.

Even when the optical combiner is the same level of transparency as
the main lens of the eyeglasses, an edge often remains at the boundary
between the two. The positions of the edges of an optical combiner are
an important factor when considering how uncomfortable the display
could be when turned off. Laterally offsetting the edges away from
PPOG can help minimize this effect.  Song and Arora conducted two
additional studies \cite{artGallaryFirst, arora2024comfortably} to
determine how far from PPOG an optical combiner's edges should be
offset.  The researchers created seven pairs of glasses with simulated
optical combiner positions ranging from 0° to -30°, with 5° offsets
first on the left (position VII from the previous study), and then
seven more in the second study from 0° to 30° to the right. The
glasses emulated optical combiners with semi-transparent
films. Participants were given these glasses and were tasked with
exploring art galleries displayed on a large computer
monitor. Participants reported feeling uncomfortable with the edge,
even though the simulated combiner had almost the same transparency as
the glasses. The first study suggested that the nearest optical
boundary to the PPOG should be offset by more than -15° (towards the
nose) from the PPOG. The second study indicated that the edges of the
optical combiner should be placed beyond -20.2° (toward the nose) or
+8.7° (toward the ear) from the PPOG horizontally. This guidance
aligns with Dowell’s \cite {dowell} 8° offset guidelines for cognitive
capture and Chua’s \cite {chua} findings that nasal positions are
uncomfortable.

{\bf Hence, right-eyed optical combiners that create obvious edges when the
  display is off should be positioned such that these edges are outside
  the range of -20.2° and +8.7° from principal position of gaze
  (PPOG). This guidance implies a virtual image with a FOV greater than 28.9°
such that the edges exceed the boundaries or an image placed
completely outside the boundaries.} 

\begin{figure}[ht]
    \centering
    \includegraphics[width=0.4\linewidth]{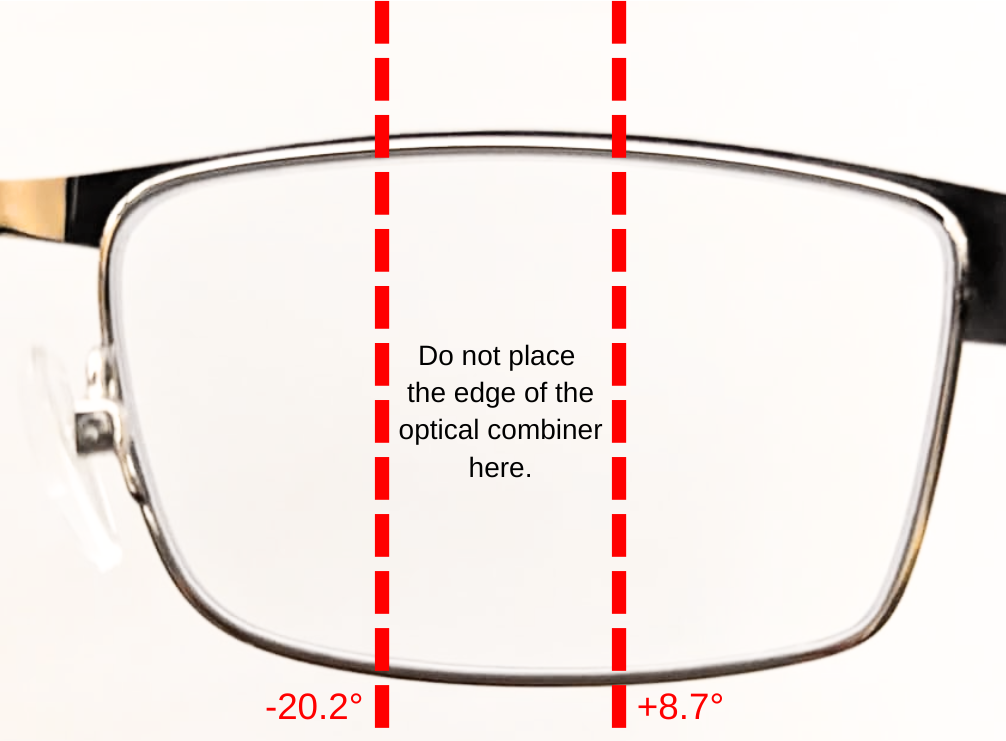}
    \caption{Avoid placing optical combiner edges within -20.2° to +8.7°}
    \label{fig:innerBounds}
\end{figure}

\subsection{Performance}

The position of the virtual image can impact a user's task
performance. Katsuyama et al.  \cite{katsuyama1989effects} studied the
effect of HUD position on user performance with object-tracking
tasks using miniature CRT displays. Katsuyama's study revealed that
performance declined with increasing azimuth (horizontal) and
elevation (vertical) angles, with upward tracking slower than downward
tracking. In the previously mentioned guided automotive maintenance
study by Zheng et al. \cite{dual_Zheng}, a line-of-sight position
yielded shorter task completion times than an above-line-of-sight
position. This result suggests that performance declines as we move
away from center.

Yang conducted similar automotive maintenance study studies
\cite{yang2015effects, yang2015study}, placing HWDs at 15° above, in,
and 15° below the PPOG. Yang found no significant differences in
overall performance among these positions. However, the
below-line-of-sight position outperformed the above-line-of-sight
position during one task step.

In the previously mentioned Chua et al. study \cite{chua}, the
center-center position had the fastest reaction time followed by
bottom-center. Post-hoc tests showed the center-center position to
have mean reaction times that showed a statistical significance
compared to top-center, top-left, and top-right.  Center-left and
center-right had similar performance to each other, though none of the
comparisons between these horizontally offset positions and the other
positions showed statistically significant differences.

In order picking studies conducted by Lin et
al. \cite{lin_towards_2021}, the bottom-right position showed a
significant decline in accuracy, perceived speed, learnability, and
comfort compared to center-center. Center-right showed a much smaller,
though statistically significant, reduction in accuracy compared to
center-center. Center-center, bottom-center and center-right positions
had a similar performance in speed.

Building on Chua's work, Rzayev et al. studied the effect of text
position on reading using a Microsoft HoloLens
\cite{Rzayev2018reading}. Rzayev compared text presentation in the
top-right, center, and bottom-center positions and found that text in
the top-right resulted in significantly lower comprehension and higher
workload compared to the center and bottom-center
positions. Additionally, reading speed was slower while walking
compared to sitting.

Overall, bottom-center and center-center perform best in the studies
presented in the literature. This advantage might be
attributed to people's greater familiarity with looking straight or
downward (as though reading a book) rather than upward \cite
{popularcould}. This hypothesis also aligns with research on upward
and downward saccadic velocity \cite {boghen1974velocity}.

{\bf In conclusion, the center-center and bottom-center positions
  perform best, and diagonally offset and top positions seem least
  favorable. The center-right position tends to perform well, though
  not as well as center-center and bottom-center.}

\subsection{Social Perception}

Another important factor in determining the position of a display is
how bystanders perceive a HWD wearer while consuming content,
especially in social settings. More specifically, if the wearer of a
HWD believes that bystanders have a negative perception of the wearer
because of their device use, this belief may hinder adoption or lead
to abandonment of the HWD. In the following discussion, we assume that
the best way to address social perception is to create a HWD that a
naive bystander would mistake for normal eyeglasses, even while the
display is being used. Note that this constraint is difficult to meet
optically. Eyeglow (i.e., the light that escapes the optical combiner
and is observable by a conversational partner) is common for many
HWDs, but there are several architectures which virtually eliminate
this issue \cite{olwal2020wearable}. Similarly, in our experience,
manual control of HWDs, using joystick rings or other devices that can
be used with the hands at the waist, is rarely noticed by naive
bystanders.  Here, though, we assume an ideal display in terms of
control and escaping light and focus instead on whether the use of the
HWD will reveal the augmentation of the wearer.

This consideration is particularly relevant for users with
disabilities who may rely on HWDs for tasks like captioning. While assistive technologies are sometimes more readily accepted by others \cite{kudlick2011black, profita2016effect, shinohara2011shadow}, some potential users avoid using them to prevent stigmatization or unwanted attention that may lead to negative social impacts \cite{kudlick2011black, profita2016effect, shinohara2011shadow}. Devices that blend into everyday items tend to be better accepted, especially by those with invisible disabilities \cite{shinohara2011shadow}. Thus, we conclude that the placement of the display should minimize visible cues to bystanders while the wearer is consuming content. While minimizing such cues  alone may not guarantee greater social acceptance \cite      {koelle2020social}, it can help reduce unwanted attention \cite{toney2003social}.

This consideration extends beyond assistive use cases to everyday
tasks. Belan et al.       \cite      {logas2021conversational}
investigated whether the change in eye gaze direction of a HWD user reveals to a
conversational partner, who is previously unaware of the display, that
the user might be reading content.  The study examined the use of a
subtle display during a face-to-face conversation with a partner who
was initially unaware of the display. Participants’ perceptions of
their conversational partner’s engagement were measured under
conditions where notifications were displayed surreptitiously at
offsets of 0, 10, and 20 degrees to the right of the HWD wearer's line
of sight, as well as a control condition with no notifications. No
significant differences in reported conversational engagement were
found between the conditions. However, once the presence of the
display was revealed, engagement scores dropped across all conditions
compared to the uninformed variant of the experiment; despite this
drop, there was still no statistically significant difference in
engagement between the control condition and the subtle display
conditions. Additionally, participants were only 40\% accurate on
average in detecting the use of the display when asked to identify
it. Interestingly, accuracy decreased when the virtual image was
offset 10 or 20 degrees versus having the virtual image centered above
the conversational partner's head.
% 0 vs. 10 and 20 degrees
In a second study comparing subtle display engagement with smartwatch engagement, participants rated a conversational partner as less engaged when using a smartwatch to monitor notifications than when using a head-up display.
% percentage?  none rated HUD worse??
In both studies, participants remained unaware of the display’s presence until it was revealed. These findings suggest that the lateral offset of a subtle display does not significantly affect its social perception. 

Another version of this study examining vertical offsets was conducted and is currently in the process of publication. We compare displays at \textit{above} line of sight (10 degrees up), line of sight, and \textit{below} line of sight (12 degrees down), against a control case. Similar to the horizontal offsetting study, no significant differences in engagement were found between the conditions, but engagement scores dropped across all conditions once participants were informed about the notifications. Informed participants were on average 38\% accurate in detecting the use of the display placed \textit{above} line of sight. Accuracy dropped to 23\% for displays placed \textit{below} line of sight, and only 3\% for displays at line of sight. Despite this, participants aware of the display still reported no significant difference in their conversation partner's engagement.

\noindent However, during verbal interviews, several female
participants expressed social discomfort during the interactions where
notifications were placed at \textit{below} line of sight, which
coincides with chest level. Further, the conversational partner who
was reading the notifications expressed discomfort as they were forced
to stare at the chest area. Based on this latter observation, we conducted another experiment in which the participant viewed notifications (instead of acting as the bystander) while a female researcher acted as their conversation partner. Participants were also told that their conversation partner did \textit{not} know they were using a display. Participants were then asked to score each position on a Likert Scale of how ``socially awkward'' the position made them feel, as well as rank the positions in order of preference. The positions included \textit{below} line of sight (-12 degrees), line of sight, \textit{above} line of sight (10 degrees), and \textit{right} (horizontally offset by 20 degrees).

\noindent Results showed 7 out of 8 male participants ranking the
\textit{below} line of sight position as least preferred, expressing
during verbal interviews that the notification at chest level was
``inappropriate'' and looking at it was ``uncomfortable.'' Both male and female participants reported a significant increase in social awkwardness for the \textit{below} position compared to all other positions. These results suggest that displays placed below eye level could discourage people from using the display during face-to-face interactions. Verbal interviews indicated two primary sources of awkwardness: the user's internal discomfort with the downward gaze, and their perception that the conversation partner might misinterpret their behavior. Even after participants were told that their conversation partner was aware of the display (and thus the reason for their gaze direction), participants did not report any reduction in awkwardness compared to when the partner was uninformed.

These results suggest that the position of the virtual image may cause
a chilling effect on the adoption and use of HWDs during social
interactions (such as captioning during face to face
conversations). Specifically, in the experiments above, {\it the               
  head-up display user's perception of their conversational partner's    perception of their eye gaze} seems to strongly effect the wearer's
comfort with the interaction. Logically, one would expect that a
HWD that made the user uncomfortable in social situations may be
quickly abandoned.

{\bf Thus, we recommend that designers avoid placing the virtual image         
  below line of sight to avoid potential social awkwardness during face to face interactions.}

\section{Resulting Position of the Display}  

From Sidenmark et al.'s \cite {sidenmark2023coordinated} work, we
expect that looking at the contents of a display using eye movement
only (as is the case with a HWD) is generally performed
within 25°. This limitation suggests that content should be restricted
to a 25° FOV if the user is expected to switch visual attention from
one side to the other or to a 50° FOV (±25°) if the user is expected
to be centering their gaze at PPOG and deflecting their gaze to one
side or the other.  Aviation safety studies recommend positioning the
display outside ±8° \cite {dowell}, as the center position is not
ideal. Additionally, Chua’s \cite {chua} work observed a preference
for right-offset displays, while left-offset positions were found to
be interruptive.

%From Sidenmark et al.'s       \cite      {sidenmark2023coordinated} work, we know that users prefer to look at displays with their eyes only within a ±25° range. Safety studies recommend positioning the display outside ±8°       \cite      {dowell}, as the center position is not ideal. Additionally, Chua’s       \cite      {chua} work observed a preference for right-offset displays, while left-offset positions were found to be distracting.

%Rzayev, Chua, and Lin's       \cite      {rzayev2020effects, chua, lin_towards_2021}, research showed that corner and top positions performed the worst. While center-aligned and bottom positions offered the best performance and comfort, they were also always in the way and interruptive. The right offset was the most preferred, striking the right balance between comfort and interruptions. Social perception studies also advised against using top and bottom positions.

Rzayev, Chua, and Lin's \cite {rzayev2020effects, chua,                        
  lin_towards_2021}, research showed that corner and top positions
performed the worst. While center-aligned and bottom positions offered
the best performance, they were also often in the way of
the user's real world task and considered interruptive. The right offset was the
most preferred, striking a balance between comfort, performance and
interruption. Social perception studies also advise against using
top and bottom positions.

To offset the display laterally, Haynes, Song, and Arora’s       \cite      {haynes2018effects, arora2024comfortably, artGallaryFirst, hwd_edge, yukunParthSherlock} work identified an optimal range of -20.2° to +8.7° for minimizing interruptions and ensuring comfort when the display is off. For extended use, they suggested positioning the display between -24.6° and +19.6°, while for shorter interactions, they recommended an extension up to 30° toward the ear. These newer findings reinforce all previous studies, confirming that displays should remain within the ±25° range for visual comfort and outside ±8° to reduce interruptions. Chua’s findings are further supported, as the left offset offers only a narrow 4.4° range for positioning before becoming interruptive.

After reviewing all the studies discussed, the ideal location for a
right-eyed display is vertically centered and extending between +8.7° and +19.6°. This results in a field of view (FOV) of 10.9°, which is smaller than the 15° FOV suggested by Martin et al. and Britain et al. \cite{britain2022preferences}. However, considering Haynes's recommendation of a 30° limit for shorter interactions and Sidenmark's 
suggested 25° limit, we can have a display from +8.7° to 23.7°, yielding an FOV of up to 15° while ensuring we remain outside of Dowell's 8° recommendation.  

{\bf Thus, we suggest offsetting a right-eyed HWD's virtual image such
  that it extends between +8.7° and +23.7° to the right of principal
  position of gaze (PPOG) and up to +30° for shorter interactions.} 

\section{Social Perception of Lens Tint}
Many factors influence whether a device is adopted, and as we've
discussed, social perception plays a key role. One element that is
often overlooked is how the design actually looks and feels to the
wearer, especially when it seems out of place or unusual in everyday
settings. Optical combiners can be visually annoying and
immediately noticeable. While developing less visible combiners
remains a technical challenge, an alternative approach is to make the
existing hardware less conspicuous. One strategy is to tint the lens
around the combiner so that they HWD looks similar to normal
sunglasses to both the wearer and any bystanders. This strategy is
used by the Magic Leap One to maintain a consistent look for the
lenses.  However, wearing tinted glasses indoors also carries social
weight. How much tint can be tolerated for everyday use HWDs?

This question was explored by Vempala et al. \cite
{vempala2024future}. Four transmissivity levels, 100\%, 75\%, 50\%,
and 25\%, were tested and participants rated how normal each
interaction felt after viewing short videos of individuals wearing the
glasses. The results showed a clear decline in perceived acceptability
as the lenses became darker, with even the 75\% transmissive lens
rated significantly lower than the fully clear one.

To refine this threshold, a second study was conducted using
finer-grain levels: 100\%, 94\%, 88\%, 82\%, and 76\%. While this
follow-up is currently under review, results suggest that
transmissivity can be reduced by up to approximately 24\% without
significantly impacting social perception. These findings provide HWD
designers with more flexibility to conceal the optical combiner without
drawing attention in everyday settings. Even slight tinting may help
blend optics into the lenses while keeping the glasses socially
acceptable.

\section{Overlaying Graphics with an Offset Display}

Some uses of augmented reality require the HWD be able to sense the
environment and register computer graphics on interest points in the
physical environment. Note that a HWD with an image offset from PPOG
is still capable of such a physically aligned overlay.  For example,
Google Glass's virtual image was placed such that the bottom-most
pixels were approximately 8° above PPOG. However, the user could
simply rotate their head downward to bring the image into alignment
with the world in front of them.  Similarly, a display offset to the
ear as suggested above can be brought into alignment with the scene in
front of the wearer by simply turning the head approximately 15° to
the left.

\section{Optics Considerations}

The angular position of a virtual image in eyewear display optics is influenced by both frame parameters and the optical architecture      \cite      {Ozan_EyewearDisplays}. We will discuss these factors separately, starting with frame parameters. For optical architectures, we will discuss freespace architectures and lightguides. 

\subsection{Frame parameters}

Two primary rigid body rotations of the lens in the eyewear frame affect virtual image positioning:

\begin{enumerate}
    \item \textbf{Pantoscopic angle:} This side-view rotation moves the virtual image vertically, either towards the ground or sky, and
    \item \textbf{Wrap angle:} This top-view rotation shifts the virtual image horizontally, towards the nose or ear of the wearer. 
\end{enumerate}

\begin{figure}[ht]
    \centering
    \includegraphics[width=0.5\linewidth]{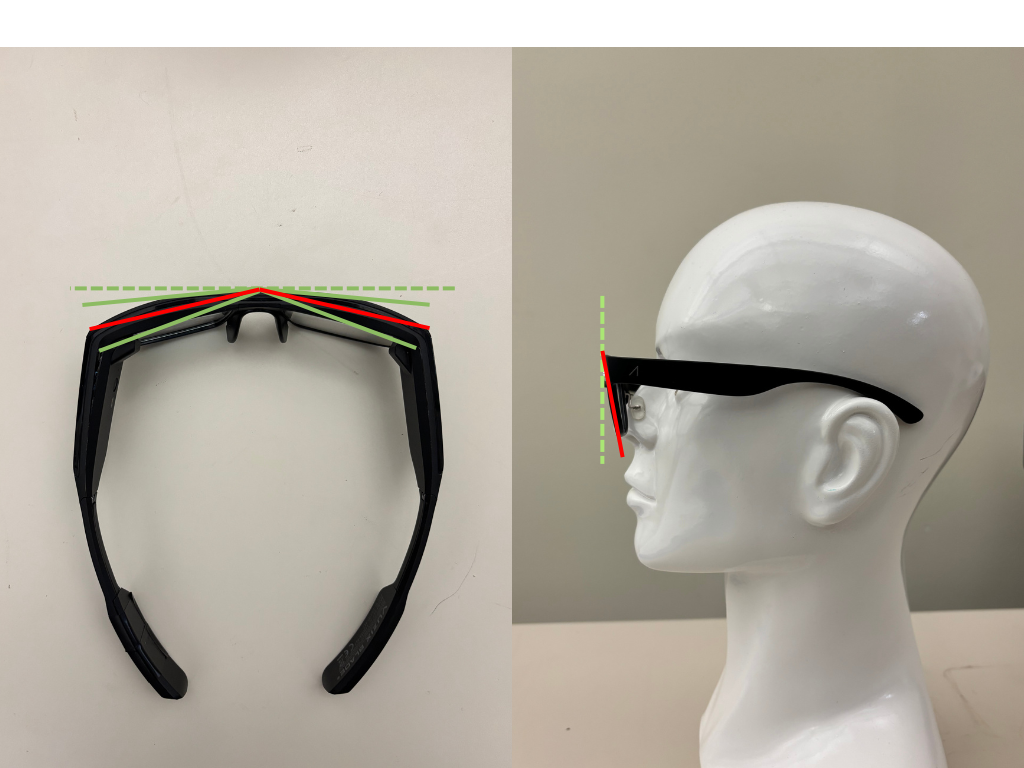}
    \caption{Eye wrap (left) and pantoscopic tilt (right).}
    \label{fig:pantascopicAndWrap}
\end{figure}

Pantoscopic and wrap angles are illustrated in Figure \ref{fig:pantascopicAndWrap}. Pantoscopic angle is added to eyewear in order to align the see-through optics with the slightly downward looking gaze of a wearer. The pantoscopic angle ranges from 5$^\circ$ to 15$^\circ$ and the wrap angle ranges from 4$^\circ$ to 11$^\circ$      \cite      {WrapPantoData}. The pantoscopic angle minimizes prism effects caused by the see-through optics. Furthermore, there are industrial design considerations of glasses positioning on the face for aesthetic and comfort reasons.

% Added rough image

If the optics were designed assuming no pantoscopic and wrap angle, once the optics are assembled onto the frame, the angular position of the virtual image will shift by the pantoscopic and wrap angle as these angles are simply rigid body rotations of the whole optics assembly. The optics and the eyewear frame must be designed jointly. Having considered the impact of frame parameters to angular virtual image position, we now turn to the influence of optical architectures, beginning with freespace designs. 

\subsection{Optical Architectures: Freespace}

Freespace architectures (e.g., Figure \ref{fig:lightGlasses}  (left)) get their name from the fact that most of the
light propagation in these systems is in air. Freespace architectures
consist of a light source (e.g., microdisplay), optics, a combiner,
and the user's pupil(s). In freespace architectures, virtual image
positioning can be obtained by a combination of tilt of the combiner
and by modifying the phase function that implements the combiner
functionality. The phase function can be implemented in hardware as a mirror, an interference pattern recorded in a holographic media, or the construction parameters (e.g. height, depth, and period; in case of a binary grating, other shapes parameterizations are also used) in a surface-relief-grating-based combiner. 
%thad: should this be "(e.g. height, depth, and period; in case of a binary grating, other shapes parameterizations are also used)"
%%% TODO: Vectorize, add labels: image source, optics, combiner, and pupil. 

\begin{figure}[ht]
    \centering
    \includegraphics[width=0.5\linewidth, trim=0 0cm 0cm 4.5cm, clip]{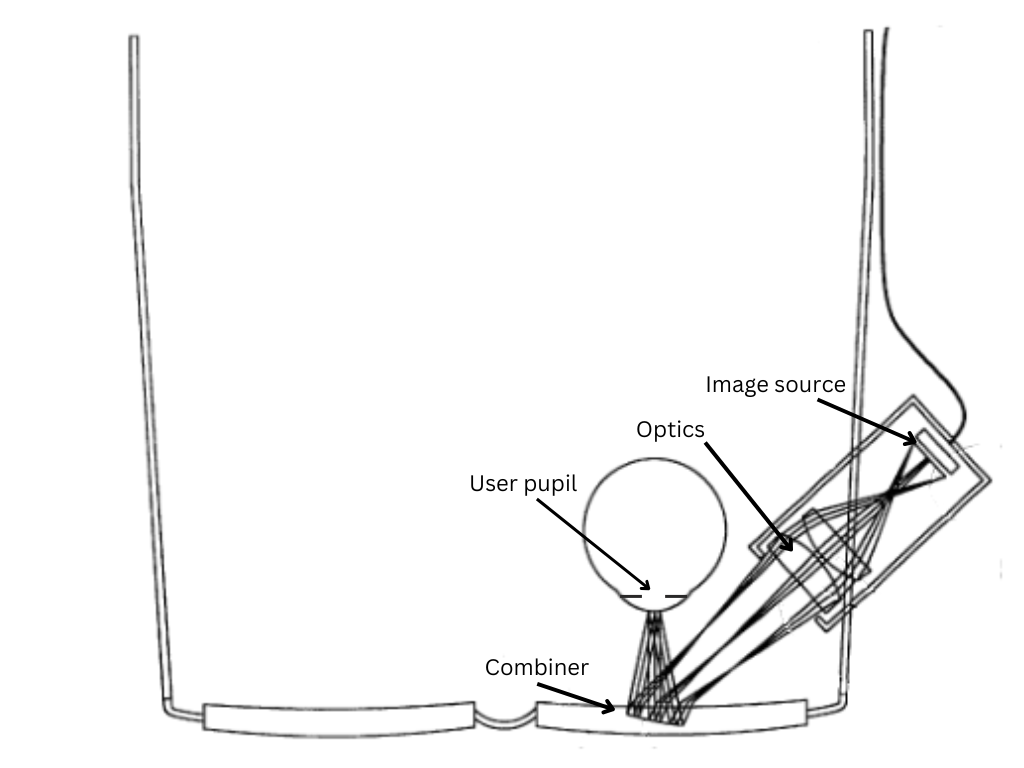}
    \caption{Example freespace architecture showing lightsource, optics, combiner, and user's pupil       \cite      {HWDReview}. Although the optics and combiners vary in the literature, these characteristic building blocks are consistent across examples. }
    \label{fig:FreespaceArchitectureExample}
\end{figure}

\subsection{Optical Architectures: Flat Lightguides}

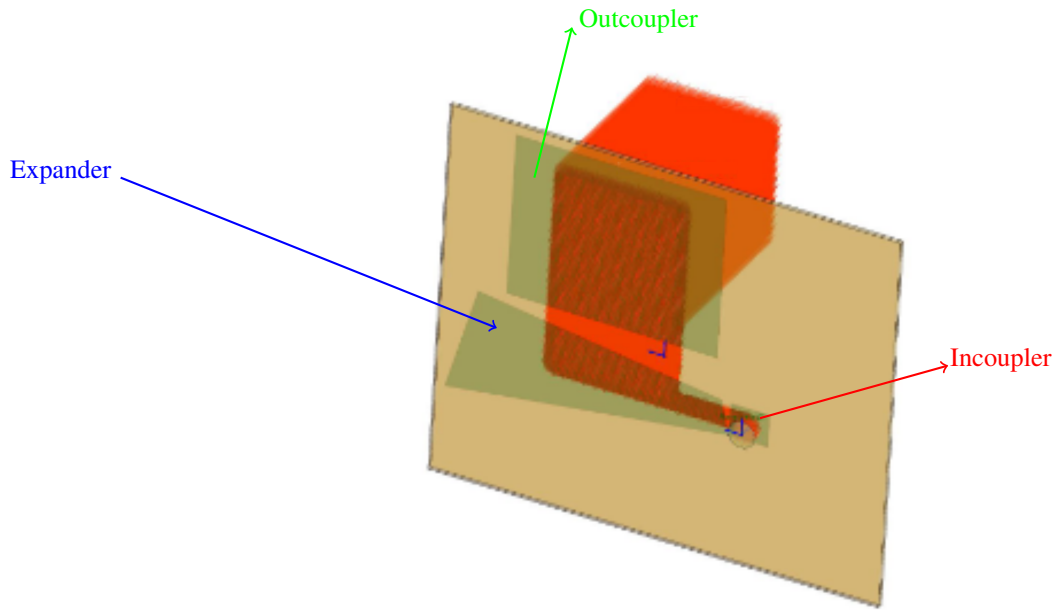
\begin{figure}[ht]
    \centering
    \begin{tikzpicture}
        \node[anchor=south west,inner sep=0] (image) at (0,0) {\includegraphics[width=0.5\linewidth, trim=0 0cm 0cm 0cm, clip]{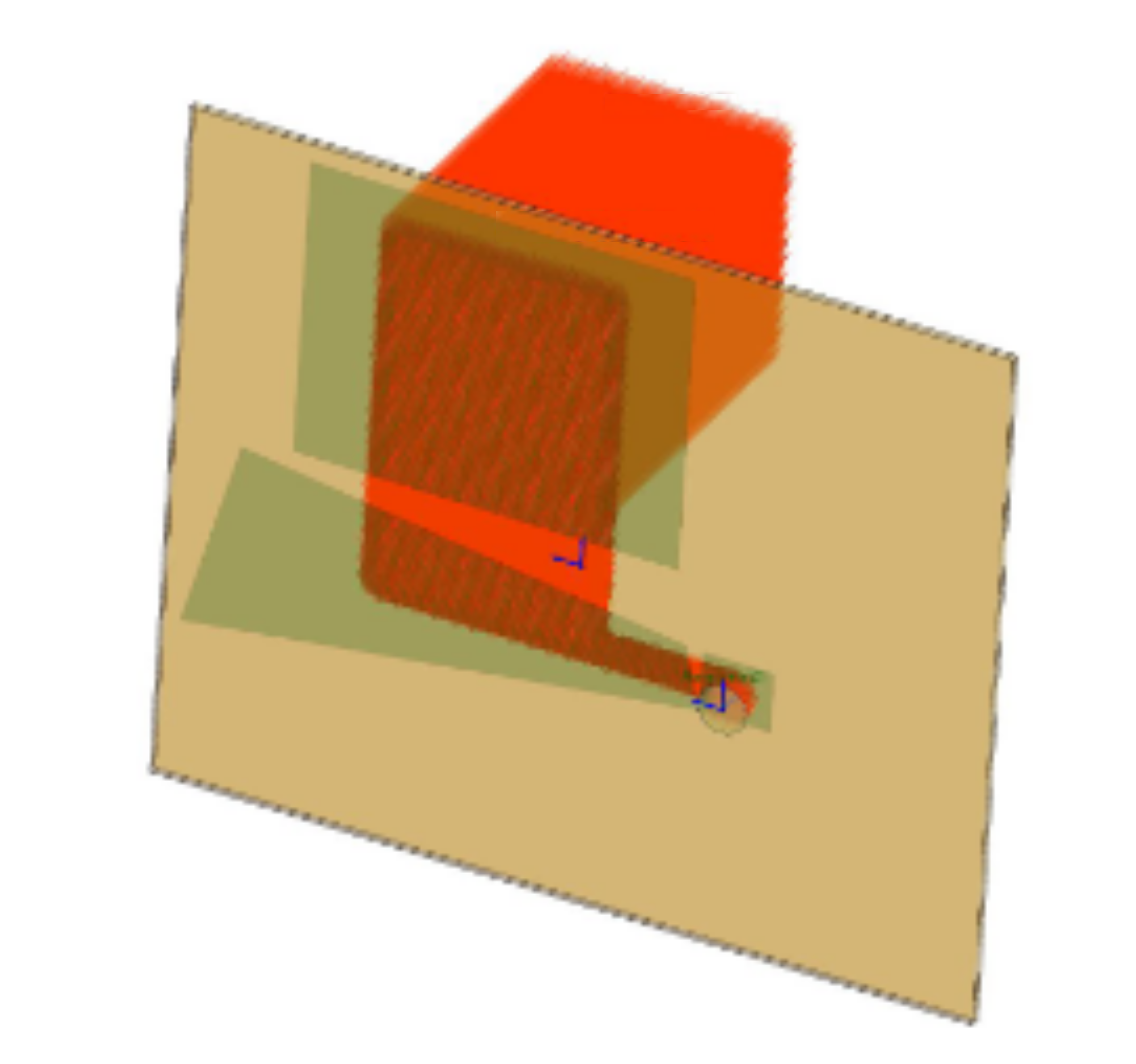}};
        % Arrows and labels
        \draw[->, thick, red] (5.5,2.8) -- (8,3.5);
        \node[above left, text = red] at (9.5, 3.3) {Incoupler};
        \draw[->, thick, blue] (-3,6) -- (2,4);
        \node[above left, text = blue] at (-3, 5.8) {Expander};
        \draw[->, thick, green] (2.5,6) -- (3,8);
        \node[above left, text = green] at (4.8, 7.8) {Outcoupler};
    \end{tikzpicture}
    \caption{Example classical lightguide architecture raytrace with an incoupler, expander, and an outcoupler. In this case, the incoupler, expander, and the outcoupler are implemented as diffraction gratings. }
    \label{fig:3Klightguide}
\end{figure}

%thad: originally this sentence included "(e.g., Fig. 1 (right))" -
%to which figure was it meant to refer?
Flat lightguide architectures (e.g., Figure \ref{fig:lightGlasses}  (right)) typically consist of a light source
(e.g., microdisplay), collimator optics, an incoupler that injects the
collimated light from the collimator optics into total internal
reflection, and an outcoupler that extracts the light from total
internal reflection towards a user's eye. A classical diffraction
grating-based lightguide with an incoupler, expander, and an
outcoupler is shown in Fig. \ref{fig:3Klightguide}. Most flat
lightguide architectures make use of 1D or 2D expansion in order to
satisfy the eyebox requirement (see Cakmakci et al. \cite{OzanEyebox} for a definition of eyebox; a full definition is out of scope in this paper). While there are a large number of flat lightguide architectures, arising mostly as an enumeration of various in/outcoupler choices, these architectures are typically designed using k-space diagrams to make sure the polychromatic field of view resides within the available angular total internal reflection bandwidth of a lightguide. For brevity, in place of using k-space diagrams, we define a lightguide as an optical element that satisfies $\vec{k}_{\text{in}} = \vec{k}_{\text{out}}$ where $\vec{k}$ is the ray/wave propagation direction with magnitude $|\vec{k}| = 2\pi/\lambda$ and $\lambda$ is the wavelength of light. The $\vec{k}_{\text{in}}$ is considered to be at the output of the collimator optics, and the $\vec{k}_{\text{out}}$ is considered to be at the output of the outcoupler. In the context of this paper, this means that any ray angle output from the collimator optics, entering the lightguide, will exit at the same angle at the outcoupler.  

We can see immediately from $\vec{k}_{\text{in}} = \vec{k}_{\text{out}}$ model of a lightguide that tilting the collimator optics, which changes $\vec{k}_{\text{in}}$, will shift the angular position of the virtual image. Tilting the collimator will also change the angle of eyeglow emission towards the worldside. Tilting the collimator could introduce artifacts in the lightguide, therefore, there is a trade-off between eyeglow emission angle, virtual image angular position, and artifacts caused by tilting the collimator optics. Collimator tilt magnitude is limited by available volume within the eyewear frame (temple or rim), creating a form factor trade-off with optics.

\subsection{Summary of Optics Considerations}

In case of freespace architectures, the optics considerations for angular position of a virtual image include the frame parameters, combiner tilt, and the  combiner phase function. In flat lightguides, optical considerations for angular position of the virtual image include the frame parameters, tilt of the collimator optics with respect to the incoupler, eyeglow considerations if applicable, and the effect of image artifacts as a function of all of these choices. In addition to these optics considerations, there are comfort, aesthetic, form factor, and industrial design considerations that must be designed jointly with the optics. 

\section{Summary of Findings for Positioning the Image of a HWD}

% Thad Slides
% \section{recs}
For a monocular right-eyed HWD, placing the image vertically centered
and between +8.7° and +23.7° degrees toward the right ear balances
performance, potential for interruption, social acceptance, and
comfort.

\begin{itemize}
    \item 
      Dowell’s aviation research suggests positioning the virtual image outside
      a 8° radius.
    \item Focusing on pixels offset >25° from PPOG may cause discomfort, keeping pixels
      <20° is best.
    \item Even for shorter interactions its recommended to keep all pixels
       <30°.
    \item Notifications displayed with a horizontal offset are less noticeable by naive bystanders observing a HWD user.
    \item Placing content below eye level in face-to-face interactions
      is socially sensitive --- both the bystander and user may feel uncomfortable when content is overlaid over the chest area.
    \item The HWD user can rotate their head to achieve visual overlay.
    \item Virtual images positioned vertically down and centered
      horizontally are often most comfortable but are also most
      interruptive and less preferred when used in the tasks tested in
      the literature.
    \item Virtual images in the corners (e.g., vertically down and
      horizontally right) and vertically up in the user's visual field
      are generally disliked and result in a decrease in task performance.
    \item For reading text, a >10° horizontal field-of-view is
      suggested; 15 degrees is a good compromise for everyday use.
\end{itemize}

% Thad, final recs?

% Hardware/Optics recs maybe?

% Several software-based optimizations can further improve the experience. 
% Sidenmark et al extensively studied interfaces using gaze for HWD that can further improve the experience       \cite      { sidenmark2023coordinated}. For hardware, optimizations can also be made, such as tinting the display to improve reading comfort, although this introduces, but alsos removes, some social perception challenges       \cite      {vempala2024future}. While we emphasize that this research focused strictly on monocular displays with a small field of view (FOV) for everyday wear, some findings may be applicable to positioning higher-FOV or binocular devices. For example, if a display has edges that may bother the user, it is crucial to make the display large enough so that both edges fall outside the range of -20.2 to 8.7 degrees, totaling approximately 28.9 degrees.

% \begin{figure}[ht]
%     \centering
%     \includegraphics[width=0.9\linewidth]{images/softwareFixes.png}
%     \caption{Gaze Based Peripheral Display from Yoshio et al.       \cite      {yoshioPeripheral} (Left and center) and Environment locked notification in simulated in VR by Rzayev et al.       \cite      {rzayev2019notification} }
%     \label{fig:softwareFixes}
% \end{figure}

\section{Conclusion}
Head worn displays (HWDs) are at the forefront of wearable technology,
with their integration into everyday life hinging on thoughtful design
decisions that balance usability, comfort, and functionality. Our
research consolidates years of findings on the optimal positioning of
displays for monocular optical see-through HWDs (OST-HWDs). Our
findings reinforce the importance of a right-offset display position
for everyday wear, with specific recommendations for placement between
+8.7° and +30° to align with human visual ergonomics and task
efficiency. For captioning-like tasks, a 15° field of view (FOV)
proves optimal, supporting content placement within the range of +8.7°
to +23.7°, while prolonged tasks may benefit from a narrower range of +8.7° to +19.6° to minimize discomfort. These guidelines are valuable for hardware manufacturers striving to design HWDs that integrate seamlessly into daily routines without compromising on performance or comfort. Future work could explore how these findings translate to other use cases and configurations, such as binocular or fully immersive systems, further refining HWD design for diverse applications.

%Bibliography
\bibliographystyle{unsrt}  
\bibliography{references}

\end{document}